\documentclass[aps,groupedaddress,floatfix,preprintnumbers,nofootinbib,eqsecnum]{revtex4}
\usepackage{amssymb,amsmath,graphicx,multirow}

\usepackage{amssymb, amsmath}
\usepackage{graphics,graphicx,epsfig,ulem}
\usepackage{float}

\usepackage{multirow}
\usepackage{longtable}
\usepackage{color}
\usepackage{xcolor}
\usepackage{slashed}

\begin{document}

{\flushright{ IPPP/16/68 \\\mbox{ } \\}}

\title{A Dynamical Mechanism for Large Volumes with Consistent Couplings}
\author{\vspace{0.2cm} Steven Abel\footnote{E-mail address:
      {\tt s.a.abel@durham.ac.uk}}\\ }
\affiliation{\vspace{0.2cm}
     IPPP, Durham University, Durham, DH1 3LE, UK }

\begin{abstract}
\vspace{0.2cm}
  {\rm
  A mechanism for addressing the ``decompactification problem'' is proposed, 
which consists of balancing the vacuum energy in Scherk-Schwarzed theories
against contributions coming from non-perturbative physics. 
Universality of threshold corrections ensures that, in such situations, the stable minimum will have consistent gauge couplings for any gauge group that shares 
the same ${\cal N}=2$ beta function for the bulk excitations as the gauge group that takes part in the minimisation.
Scherk-Schwarz compactification from 6D to 4D in heterotic strings is discussed explicitly, together with two alternative possibilities for the non-perturbative physics, namely metastable SQCD vacua and a 
single gaugino condensate. In the former case, it is shown that modular symmetries gives various consistency checks, and allow one to follow soft-terms, playing a similar role to $R$-symmetry in global SQCD. The latter case is
particularly attractive when there is nett Bose-Fermi degeneracy in the massless sector. In such cases, because the original Casimir energy is generated 
entirely by excited and/or non-physical string modes, it is completely immune to the non-perturbative IR physics.
The separation between UV and IR contributions to the potential greatly simplifies the analysis of stabilisation, and is 
a general possibility that has not been considered before.
}
\end{abstract}

\maketitle

\def\beq{\begin{equation}}
\def\eeq{\end{equation}}
\def\beqn{\begin{eqnarray}}
\def\eeqn{\end{eqnarray}}
\def\half{{\textstyle{1\over 2}}}
\def\quarter{{\textstyle{1\over 4}}}

\def\calO{{\cal O}}
\def\calC{{\cal C}}
\def\calE{{\cal E}}
\def\calT{{\cal T}}
\def\calM{{\cal M}}
\def\calN{{\cal N}}
\def\calF{{\cal F}}
\def\calS{{\cal S}}
\def\calY{{\cal Y}}
\def\calV{{\cal V}}
\def\ibar{{\overline{\imath}}}
\def\chibar{{\overline{\chi}}}
\def\ttwo{{\vartheta_2}}
\def\tthree{{\vartheta_3}}
\def\tfour{{\vartheta_4}}
\def\ttwob{{\overline{\vartheta}_2}}
\def\tthreeb{{\overline{\vartheta}_3}}
\def\tfourb{{\overline{\vartheta}_4}}
\def\Str{{{\rm Str}\,}}

\def\bfell{{\boldsymbol \ell}}
\def\xx{\hspace{0.3cm}}
\def\xxl{\hspace{0.285cm}}
\def\xxh{\hspace{0.237cm}}
\def\yy{\hspace{0.115cm}}
\def\yyr{\hspace{-0.02cm}}

\def\qbar{{\overline{q}}}
\def\mm{{\tilde m}}
\def\nn{{\tilde n}}
\def\rep#1{{\bf {#1}}}
\def\ie{{\it i.e.}\/}
\def\eg{{\it e.g.}\/}

\newcommand{\newc}{\newcommand}
\newc{\gsim}{\lower.7ex\hbox{$\;\stackrel{\textstyle>}{\sim}\;$}}
\newc{\lsim}{\lower.7ex\hbox{$\;\stackrel{\textstyle<}{\sim}\;$}}

\newcommand{\red}[1]{\textcolor{red}{#1}}

\hyphenation{su-per-sym-met-ric non-su-per-sym-met-ric}
\hyphenation{space-time-super-sym-met-ric}
\hyphenation{mod-u-lar mod-u-lar--in-var-i-ant}


\def\inbar{\,\vrule height1.5ex width.4pt depth0pt}

\def\IC{\relax\hbox{$\inbar\kern-.3em{\rm C}$}}
\def\IQ{\relax\hbox{$\inbar\kern-.3em{\rm Q}$}}
\def\IR{\relax{\rm I\kern-.18em R}}
 \font\cmss=cmss10 \font\cmsss=cmss10 at 7pt
\def\IZ{\relax\ifmmode\mathchoice
 {\hbox{\cmss Z\kern-.4em Z}}{\hbox{\cmss Z\kern-.4em Z}}
 {\lower.9pt\hbox{\cmsss Z\kern-.4em Z}} {\lower1.2pt\hbox{\cmsss
 Z\kern-.4em Z}}\else{\cmss Z\kern-.4em Z}\fi}

\long\def\@caption#1[#2]#3{\par\addcontentsline{\csname
  ext@#1\endcsname}{#1}{\protect\numberline{\csname
  the#1\endcsname}{\ignorespaces #2}}\begingroup \small
  \@parboxrestore \@makecaption{\csname
  fnum@#1\endcsname}{\ignorespaces #3}\par \endgroup}
\catcode`@=12

\input epsf


\tableofcontents

\section{Introduction}

The Scherk-Schwarz (SS) mechanism is one of the most
attractive means of spontaneously breaking supersymmetry
(SSSB) \cite{Scherk:1978ta,Scherk:1979zr}.
In the SS mechanism, supersymmetry is broken by compactification with $R$-symmetry violating
boundary conditions, and from a 4D perspective the inverse volume $1/R^{d}$
(where $R$ is used as a generic compactification scale) plays the
role of an order parameter for supersymmetry breaking in the effective
field theory. This yields all-orders control over supersymmetry breaking,
and shields dimensionful operators
such as the Casimir energy and soft-terms from the ultra-violet (UV) completion \cite{Antoniadis:1990ew, Antoniadis:1991fh, Antoniadis:1998sd,Ghilencea:2001bv}.
They can then largely be computed as finite Kaluza-Klein (KK) contributions
in an effective extra-dimensional field theory, enhancing predictivity.
There are numerous interesting phenomenological applications, for
example in the recent work of refs.~\cite{Dimopoulos:2014aua,Cohen:2015gaa,Garcia:2015sfa,Tobioka:2015vsv,Antoniadis:2015chx,Reece:2015qbf}. 

In such theories, a volume significantly larger than the fundamental scale, $R\gg \ell_{s}$, is necessary (even if one does not
insist on low scale supersymmetry breaking) if one wishes the reproduce the physics of the traditional field theory SS mechanism. This is because 
 heavy modes come to dominate over the KK modes in loop processes
once the compactification radius approaches the fundamental length
scale (see the discussion in ref.~\cite{Abel:2015oxa}). In the context of non-supersymmetric
string theory for example, ``non-physical''
proto-gravitons start to be important once $R\lesssim 2 \ell_s$.
The necessary separation between the UV completion and the KK scale 
can be achieved by configurations that
interpolate from supersymmetric theories at large radius to non-supersymmetric
ones at small radius \cite{Abel:2015oxa}. Ideally, one would then like to treat this as an approximate 
``moduli space'', and generate a consistent supersymmetry
breaking solution at large volume dynamically. This has been widely discussed
in the Scherk-Schwarz context in for example refs.~\cite{ponton,Borunda:2002ra,vonGersdorff:2003rq,Dudas:2004vi,Dudas:2005vna,
vonGersdorff:2005ce,Angelantonj:2006ut,Braun:2007ff,Gross:2008he}. 

However large volumes are problematic in the context of heterotic string theory. They are felt universally by the gauge
couplings, which are then generally rendered inconsistent at one-loop by the corresponding KK mode contributions. This is a generic source of tension for the SS mechanism in heterotic strings and indeed any SS set-up that does not have a ``brane'' configuration. 

To be specific,
consider an effective 5D SQCD theory  (i.e. one in which only one
compactified dimension is significantly larger than the fundamental
scale). Supposing that any other moduli except the radius are already
stabilised at small volumes (so they play no further role in the dynamics
or in the magnitude of the gauge couplings) the expression
for the gauge coupling of the effective 4D SQCD theory is 
\begin{equation}
\frac{16\pi^{2}}{g^{2}(\mu)}=k\frac{16\pi^{2}}{g_{s}^{2}}+b\ln\frac{M_{s}^{2}}{\mu^{2}}+\Delta(R)\, ,\label{eq:flow}
\end{equation}
where $b$ is the beta function coefficient of the original
effective 4D $\mathcal{N}=1$ theory (in a convention where $SU(N)$
supersymmetric QCD with $F$ flavours would have $b=-3N+F$), and $\Delta$
are the offending threshold contributions which at large volumes are dominated by the KK
sector of the theory, 
\begin{equation}
\Delta(R)=CRM_{s}-2b\ln(RM_{s})\, .
\end{equation}
The constant $C$ depends on various other parameters and moduli,
most importantly on the beta functions of the $\mathcal{N}=2$
content of the theory. In this preliminary discussion (and in fact right up to the last section)  $g_s$  
will be assumed to be fixed beforehand: ultimately
though it will also be dynamical, being given by the VEV of the axio-dilaton. 

There are then two possibilities assuming that $C\neq0$. Gauge
couplings that have $C>0$ are made weaker by the threshold
corrections. Broadly speaking one can interpret this as the contribution
from power-law running between the fundamental scale and the KK scale
\cite{Dienes:1998vh,Dienes:2002bg} (although there are various subtleties
in mapping extra-dimensional field-theory to string theory -- see for
example ref.~\cite{Ghilencea:2001bv}). At large volume the couplings
become tiny and the corresponding symmetry is to all intents and purposes
global. By contrast those couplings that have $C<0$ grow stronger
at large radius, from extremely weak values 
at the fundamental scale. They can in principle become reasonably
large, but then one has to balance
the threshold contribution to $1/g^{2}$ against its tree-level value.
It should be noted that $C$ and $b$ need not have the
same sign, so there is nothing to prevent a theory flowing to stronger coupling
at the KK scale, and then for the effective 4D theory to be IR-free (and vice-versa); 
to simplify the discussion it will be assumed that they do have
the same sign.

To summarise the difficulty, $C>0$ couplings are insignificant at low energy
unless the gauge symmetry is localized in the large volume, implying some kind
of brane set-up. On the other hand, $C<0$ couplings seem to
imply a fine-tuning of tree-level against radiative corrections, so
that they are extremely weak at the string scale, but order one just
at the bottom of the KK tower where they enter the logarithmically
running 4D regime. This issue, which has become
known as the ``decompactification problem'', has been discussed in the past in for
example in refs.~\cite{Caceres:1996is,Kiritsis:1996xd,Kiritsis:1998en,Antoniadis:2000vd,Dienes:2002bg,Faraggi:2014eoa,Kounnas:2015yrc,Kounnas:2016gmz}, and was eloquently summarized recently in ref.~\cite{Partouche:2016xqu}. Special
theories are known that circumvent the coupling/volume sensitivity
because they {\it do} have $C=0$ \cite{Faraggi:2014eoa,Kounnas:2015yrc,Kounnas:2016gmz,Partouche:2016xqu},
but here it will be of interest to consider more generic models.  

The purpose of this paper is to argue that there {is} in fact a way to realise
order one couplings at large volume  {\it dynamically and without fine-tuning}, providing
a solution to the decompactification problem for a much broader class
of models. The set-up is very general: it requires only that the compact volume 
is stabilised by balancing a dynamical transmutation scale, $\Lambda_{e}$, against a leading order one-loop 
Casimir energy. This results in a gauge coupling that is inevitably becoming large
precisely where the volume is stabilised. The particular gauge factor
that takes part in the stabilisation may of course be of little further use
for phenomenology, depending on the precise non-perturbative physics behind the appearance of $\Lambda_e$. However the universality 
in the gauge couplings and their ${\cal N}=2$ threshold corrections  
ensures that {\it any gauge group with the same }$C$ {\it will
also have gauge couplings of order one}, with only logarithmic differences
appearing due to the different $\mathcal{N}=1$ beta functions, $b$.
(Note the gauge group and particle content do not have to be the same, 
so for example the content of an ${\cal N}=2$  $SU(5)$ SQCD with 6 flavours has a  $C$ equal to that of 
${\cal N}=2$ $SU(3)$ SQCD with 2 flavours.) 
That such universality exists even in theories that have supersymmetry
broken by the Scherk-Schwarz mechanism has been recently shown in
ref.~\cite{Angelantonj:2015nfa}. Meanwhile those gauge factors with
larger or smaller $C$ will become effectively global or strongly coupled and confined,
respectively, and will play little further role in phenomenology. 

The configuration that will be studied here is based on the
interplay of two competing mildly repulsive and mildly attractive
effects. The first is the aforementioned Casimir energy that arises in
compactifications where supersymmetry is spontaneously broken by the
SS mechanism. This typically goes as $(N_{f}^{0}-N_{b}^{0})/R^{4}$,
where $R$ is the compactification scale along the direction that
breaks supersymmetry, and $(N_{f}^{0}-N_{b}^{0})$ is the nett Fermi-Bose
number of the states left massless by the SS mechanism; choosing
it to be positive, it represents a repulsive effect running away to
large radius. The competing effect is a positive contribution to
the cosmological constant arising from some non-perturbative process. We will consider  
two options: the first is an SQCD sub-sector of the theory
which sits in the metastable supersymmetry breaking minimum of Intriligator,
Seiberg and Shih \cite{Intriligator:2006dd} (ISS) and the second is a Yang-Mills gaugino condensate. 
Both of these produce terms that are governed by the dynamical scale of the theory, which in turn depends
on the threshold contribution to the effective gauge coupling in eq.(\ref{eq:flow}).
Assuming that both $C_{}\mbox{ and }b_{}$ are negative, this
contribution increases with radius, so it is attractive. 

The result is that the theory is driven dynamically to the boundary of the {{perturbative}} 
moduli space and minimised there, with all gauge couplings that share the same value of $C$ automatically taking values of order one
no matter how small the (universal) string-scale value. It is clear that the resulting large volume is then directly related to the 
smallness of the string-scale coupling at the origin. 

The next section presents a 5D toy-version of the mechanism, expressed purely in field theory. 
It emphasises the general difference between an SS vacuum energy that is broadly the same as the field theoretical one described above, and the qualitatively different possibility that heavy UV modes in the theory dominate the SS vacuum energy. This may simply be a result of the 
volume approaching the string scale, in which case (as mentioned above) the leading contributions come from non-physical modes, or it may be a result of the massless contributions vanishing in theories that have $(N_{f}^{0}=N_{b}^{0})$, in which case the leading contributions come from the lowest lying string excitations. In these cases the SS vacuum energy cannot be well understood in extra-dimensional field theory, but can be easily calculated in string theory. Moreover an important and recurring theme is that, because it is UV in nature, the SS vacuum energy in such cases is completely immune to any non-perturbative physics that one might balance it against in order to produce  a stable compactification. In order to emphasise the distinction, this kind of SS induced vacuum energy will be referred to as {\it {UV-Casimir energy}}.

Section \ref{seciii}  collects the necessary ingredients required for the  string realisation. One of the reasons for interest in the ISS mechanism in this context
rather than just gaugino condensation will become clear: it allows several checks of the stringy implementation of non-perturbative supersymmetry breaking, and in the generic SS case it gives a cleaner separation between the contributions to the potential coming from the SS and ISS mechanisms. The Casimir energy is calculated in toroidal SS compactifications from 6D to 4D, the residual modular symmetry is discussed and several new results are presented, on the use of modular invariance to follow the SS induced soft terms, and on a consistency condition for the stringy implementation of the ISS mechanism.

These results are used Section \ref{seciv} to study stabilisation for generic Casimir energies, and also for the case in which an exponentially suppressed UV-Casimir balances against a gaugino condensate. 
Up to this point, the approach is somewhat modular  in that the 
tree-level coupling $g_s$ and also its axionic partner are taken to be fixed parameters in order to investigate how the compactification dynamics adjusts to 
consistently accommodate tiny values. In this last example all moduli ($S,T,U$) are treated as dynamical fields. The beauty of UV-Casimir energy becomes evident here, and it is worth repeating it: because it is blind to IR physics, one can essentially balance two robustly independent contributions to the vacuum energy that are nevertheless functions of only the three $S,T,U$ moduli. An additional interesting feature here is that the gaugino condensate scale automatically adjusts to roughly match that of the UV-Casimir energy.

\section{The mechanism in a 5D non-maximal Scherk-Schwarz model}
\label{section2}
It is convenient to proceed by developing the 5D example of the mechanism outlined in the
Introduction, with the non-perturbative physics being the ISS mechanism. Although it illustrates the principle, it should 
be regarded as something of a warm-up
exercise to the more stringy implementation in forthcoming sections.
In particular, an important question is whether the soft-terms induced
by the SS mechanism can disrupt the supersymmetry breaking of the
ISS mechanism, which is after all written entirely within $\mathcal{N}=1$
supersymmetric QCD. In the next section, we shall learn how to treat
this question by mapping soft-terms using the modular symmetry of
the $6D\rightarrow4D$ compactification. There we will also consider gaugino condensation as an 
alternative non-perturbative mechanism. For the moment we shall solve 
this issue by invoking non-maximal SS phases. 

It will be sufficient
to assume that the Scherk-Schwarz action shifts the masses of vector-like
pairs of states. (It could also act on chiral states but it would
not qualitatively change the discussion.) The KK masses take the form
$(n+q_{F\pm})/R$, and $(n+q_{B\pm})/R$, where $q_{B\pm}=(\pm\alpha_{B}+Rm_D)$
and $q_{F\pm}=(\pm\alpha_{F}+Rm_D)$, and where $m_D$ is an elementary
supersymmetric Dirac mass (a.k.a. $\mu$-term). 

There are limitations as to where the mechanism can work in its most
naive form. As mentioned above the main constraint arises from the
fact that the results of ISS are derived in 4D $\mathcal{N}=1$ supersymmetric
QCD, whereas this is a 5D setting in which supersymmetry is already
partially broken by the SS mechanism. If one wishes to adopt the ISS
results at face-value (with no extra KK modes to complicate things), one can impose a modest energy gap between
the dynamical scale of the the SQCD theory and the mass-scale of the
lowest lying KK modes, and in addition between the two sources of
superymmetry breaking to ensure that the ISS analysis is not disrupted
by the soft-terms that are already induced by the SS mechanism. The
latter are expected to remain of order $\alpha_{F,B}/R$ throughout
(in both the electric and magnetic SQCD phases), so the ISS results
can be used wholesale if this scale is much less than the supersymmetry
breaking induced in the low energy theory of the ISS mechanism. This
can be achieved by assuming non-maximal Scherk-Schwarz phases, $\alpha_{F,B}\ll1/2$.
Such non-maximal phases are somewhat artificial in the stringy Scherk-Schwarz mechanism 
\cite{Rohm, Ferrara:1987es, Ferrara:1987qp, Ferrara:1988jx, Kounnas:1989dk, Kiritsis:1997ca, Dudas:2000ff,Scrucca:2001ni, Borunda:2002ra, Angelantonj:2006ut}
because $\alpha_{F,B}$ are proportional to some linear combination
of gauge and $R$-charges and can only take discrete values. In some
orbifold compactifications, these could be for example $1/5$, but
they cannot be arbitrarily small. As mentioned, a more realistic implementation
will ultimately require a proper treatment of the mapping of soft-terms
in the SS context, including KK modes, and a properly adjusted ISS picture to take account
of them. 

The last constraint is on the elementary supersymmetric Dirac mass
required in the ISS mechanism: it should take values $m_D\ll1/R$.
It is simple and natural \textendash{} although not crucial \textendash{}
to take $m_D$ also to be induced by the compactification, so that
it too is proportional to $1/R$, with constant of proportionality
$\alpha_{D}=Rm_D\ll1$. In this 5D model therefore, we shall maintain
the following hierarchy of scales:
\begin{equation}
\frac{1}{R}\gtrsim\Lambda_{e}\gg\sqrt{\Lambda_{e}\alpha_{D}/R}\gg\frac{\alpha_{F,B}}{R}\,.\label{eq:hierarchy}
\end{equation}
The left-most scale is the bottom of the KK tower, which is taken to be greater
than the dynamical scale $\Lambda_{e}(R)$ of the effective 4D SQCD
theory. Meanwhile $m_D$ must be smaller than $\Lambda_{e}(R)$
so that states which get a Dirac mass are not simply integrated out.
And finally, on the right, a {\it sufficient condition} for the 4D
$\mathcal{N}=1$ ISS analysis to be a good approximation, is that
the scale of effective supersymmetry breaking induced by the Scherk-Schwarz
mechanism is negligible compared to the supersymmetry breaking induced
later by the ISS mechanism. These constraints translate into a condition
on $\Lambda_{e}R$ of
\begin{equation}
1\gtrsim R\Lambda_{e}\gg\alpha_{D}\,,\,\frac{\alpha_{F,B}^{2}}{\alpha_{D}}\,.\label{eq:hierarchy2}
\end{equation}
It will be convenient to assume $\alpha_{D}\sim\alpha_{F,B}$. 

\subsection{The generic Casimir energy case}
The potential may now be determined, beginning with the Casimir contribution.
For definiteness let us take $N_{b}^{0}$ of the $\alpha_{B}$ and
$N_{f}^{0}$ of the $\alpha_{F}$ to be exactly zero, and the rest
to be degenerate with $\alpha_{B}=\alpha_{F}=\alpha\ll1$. The light
theory then has $N_{f}^{0}$ massless fermions and $N_{b}^{0}$ massless
bosons, with the remainder having mass $\sim\alpha/R$. The one-loop
Casimir energy can be computed at the level of the 5D KK
theory regardless of any more fundamental UV completion, because it
is dominated by the massless modes and their KK excitations (assuming
that the KK levels do not have equal numbers of fermions and bosons).
The simplest method is to Poisson resum the Schwinger integral form
of the Coleman-Weinberg potential;
\begin{eqnarray}
V_{C} & = & -\frac{1}{16\pi^{2}}\mbox{Tr}\sum_{n}\int_{0}^{\infty}\frac{dt}{t^{3}}\,\exp\left[-t(n+q_{B+})^{2}/R^{2}\right]+\exp\left[-t(n+q_{B-})^{2}/R^{2}\right]\nonumber \\
 &  & \,\,\,\,\,\qquad\,\,\,\,\,\,\,\,\,\,\,\,\,\qquad-\exp\left[-t(n+q_{F+})^{2}/R^{2}\right]-\exp\left[-t(n+q_{F-})^{2}/R^{2}\right]\,,
\end{eqnarray}
where the trace is over the supermultiplet representations. The insensitivity
of the Casimir energy to the UV-completion is evident here in the
fact that there is no need for a UV cut-off on the integral. (In other
words a full string calculation as in ref.~\cite{Abel:2015oxa} would just
give additional exponentially suppressed corrections.) Poisson resumming
this expression gives
\begin{eqnarray}
V_{C} & = & -\frac{1}{16\pi^{2}}\mbox{Tr}\int_{0}^{\infty}dtR\pi^{1/2}t^{-7/2}\sum_{\ell=-\infty}^{\infty}e^{-\ell^{2}\pi^{2}R^{2}/t}[\cos(2\pi\ell q_{B+})+\cos(2\pi\ell q_{B-})\nonumber \\
 &  & \qquad\qquad\qquad\qquad\qquad\qquad\qquad\qquad\qquad-\cos(2\pi\ell q_{F+})-\cos(2\pi\ell q_{F-})]\,\,,
\end{eqnarray}
and performing the integral gives 
\begin{equation}
V_{C}=\mbox{Tr}\left[B(q_{F+})+B(q_{F-})-B(q_{B+})-B(q_{B-})\right]\,,
\end{equation}
where (in agreement with e.g. \cite{Antoniadis:1998sd,Ghilencea:2001bv,Borunda:2002ra,Cohen:2015gaa,Tobioka:2015vsv})
\begin{equation}
B(x)=\frac{3}{64\pi^{6}R^{4}}\left(\mbox{Li}_{5}e^{2\pi ix}+\mbox{Li}_{5}e^{-2\pi ix}\right)\,.
\end{equation}
Expanding in the $\alpha$'s gives, 
\begin{equation}
V_{C}=\frac{3\zeta(3)}{8\pi^{4}}\frac{(N_{f}^{0}-N_{b}^{0})\alpha^{2}}{R^{4}}\,.
\end{equation}

The second ingredient for the potential is of course the ISS contribution
from an $SQCD$ sector. Assuming that the original theory contains
an $SU(N)$ gauge group with $F$ flavours of fundamental/antifundamental
pairs of chiral superfields, the potential comes from the O'Raighfeartaigh
superpotential of the magnetic SQCD theory, and takes the form 
\begin{equation}
W_{ISS}=h\mbox{Tr}(q\Phi\tilde{q})-\frac{\alpha_{D}\Lambda_{e}}{R}\mbox{Tr}(\Phi)\,,\label{eq:ISS}
\end{equation}
where $q,\,\tilde{q}$ are magnetic quarks, $\Phi$ is the $F\times F$
bound state meson, and where ignorance about the precise normalization
of $\Phi$ has been absorbed into the parameters $\alpha_{D}$ and
$h$ \footnote{More precisely, following ref.\cite{Intriligator:2006dd}, if
the original SQCD theory has a dynamical scale $\Lambda_{e}$, a superpotential
$W_{e}=m_{D}Q\tilde{Q}$, and a canonically normalized meson $\hat{\Phi}=\gamma^{-1}Q\tilde{Q}/\Lambda_{e}$,
then $W_{ISS}\equiv\sqrt{\gamma}h\mbox{Tr}(q\hat{\Phi}\tilde{q})-\sqrt{\gamma}m_{D}\Lambda_{e}\mbox{Tr}(\hat{\Phi})$,
with the understanding that $W_{ISS}$ is to be treated as a global
superpotential. This issue will become important later and will be
revisited, when a proper distinction between the physical and holomorphic
scales will be made.}.

Provided that the number of colours and flavours is such that the
SQCD theory is in the free magnetic window, $N+1<F\leq3N/2$, the
result is an additional tree-level term in the potential of the form
\begin{equation}
V_{ISS}=N\alpha_{D}^{2}\left(\frac{\Lambda_{e}}{R}\right)^{2}.
\end{equation}
The total potential is 
\begin{equation}
V=R^{-4}\left[\alpha^{2}\rho (N_{f}^{0}-N_{b}^{0})+\alpha_{D}^{2}N\left(R\Lambda_{e}\right)^{2}\right]\,,\label{eq:potential}
\end{equation}
where $\rho=\frac{3\zeta(3)}{8\pi^{4}}\approx5\times10^{-3}\ll1$. 

As an aside, note that for {\it negative} Casimir
energy the potential can be precisely zero while still satisfying
the conditions in eq.(\ref{eq:hierarchy2}) for the $\mathcal{N}=1$
supersymmetric ISS analysis to be valid: indeed a zero potential requires
only 
\begin{equation}
\left(R\Lambda_{e}\right)^{2}=\frac{(N_{b}^{0}-N_{f}^{0})}{N}\rho \frac{\alpha^{2}}{\alpha_{D}^{2}}\,.
\end{equation}

Conversely, for positive Casimir energy, one may now invoke the $R$-dependence
of $\Lambda_{e}$ using eq.(\ref{eq:flow}) to minimize the potential.
Taking $k=1$ and setting the dynamical scale
of the $SQCD$ theory to be where $1/g^{2}(\Lambda_{e})=0$
gives 
\begin{equation}
\left(R\Lambda_{e}\right)^{2}=e^{-\frac{16\pi^{2}}{g_{s}^{2}|b_{}|}+\frac{C_{}}{b_{}}RM_{s}}\,.\label{eq:rlam}
\end{equation}
It is convenient to define a fiducial coupling $g_{0}$ (which is
of order $g_{s}$), and a corresponding fiducial scale, $\mu_{0}$,
given by
\begin{eqnarray}
\frac{16\pi^{2}}{g_{0}^{2}} & = & \frac{16\pi^{2}}{g_{s}^{2}}+b_{}\ln\frac{M_{s}^{2}}{\mu_{0}^{2}},\nonumber \\
\frac{\mu_{0}^{2}}{M_{s}^{2}} & = & \frac{\alpha^{2}\rho (N_{f}^{0}-N_{b}^{0}) }{\alpha_{D}^{2}{N}} \sim10^{-2}\,.\label{eq:g2}
\end{eqnarray}
The full potential has a minimum at 
\begin{eqnarray}
\label{eq:mini}
R_{min}M_{s} & = & \frac{b_{}}{C_{}}\left[4+W\left(4e^{-4}\frac{\mu_{0}^{2}}{M_{s}^{2}}e^{\frac{16\pi^{2}}{|b_{}|g_{s}^{2}}}\right)\right]\nonumber \\
 & = & \frac{1}{|C_{}|}\frac{16\pi^{2}}{g_{0}^{2}}+\mathcal{O}(1)\,,
\end{eqnarray}
where $W$ is the Lambert $W$-function (a.k.a. product log). 
Eq.(\ref{eq:g2}) then gives 
\begin{equation}
R_{min}\Lambda_{e}=\mu_{0}/M_{s}.
\end{equation}

If the parameters are all of similar magnitude, $(N_{f}^{0}-N_{b}^{0})/N_{c}\sim\alpha_{D}/\alpha\sim1$,
then eq.(\ref{eq:g2}) gives $R_{min}\Lambda_{e}\sim\sqrt{C}\approx0.07$,
automatically satisfying the requirement in eq.(\ref{eq:hierarchy2})
and achieving the desired effect of the QCD theory ending up with
a dynamical scale somewhat below the KK mass-scale, $M_{KK}=1/R_{min}$,
even if $16\pi^{2}/g_{s}^{2}\sim R_{min}M_{s}$ is chosen to be huge.
In order to satisfy the other constraints of eq.(\ref{eq:hierarchy2}),
under the assumption that $\alpha_{D}\sim\alpha_{F,B}=\alpha$ one
requires only that $\alpha^{2}\ll \rho\frac{(N_{f}^{0}-N_{b}^{0})}{N}$
which is relatively easy to achieve. (For example with $N=N_{f}^{0}-N_{b}^{0}$,
one requires $\alpha\lesssim1/10$ which is conceivably possible even within some string orbifold models). The value of the cosmological
constant at the minimum is given by 
\begin{eqnarray}
V(R_{min}) & = & R_{min}^{-4}\left[2\rho \alpha^{2}(N_{f}^{0}-N_{b}^{0})\right]\,\nonumber \\
 & \approx & \frac{g_{s}^{8}M_{s}^{4}}{(16\pi^{2})^{4}}\left[2\rho \alpha^{2}(N_{f}^{0}-N_{b}^{0})\right].
\end{eqnarray}

\begin{figure}
\noindent \centering{}
\includegraphics[scale=0.3]{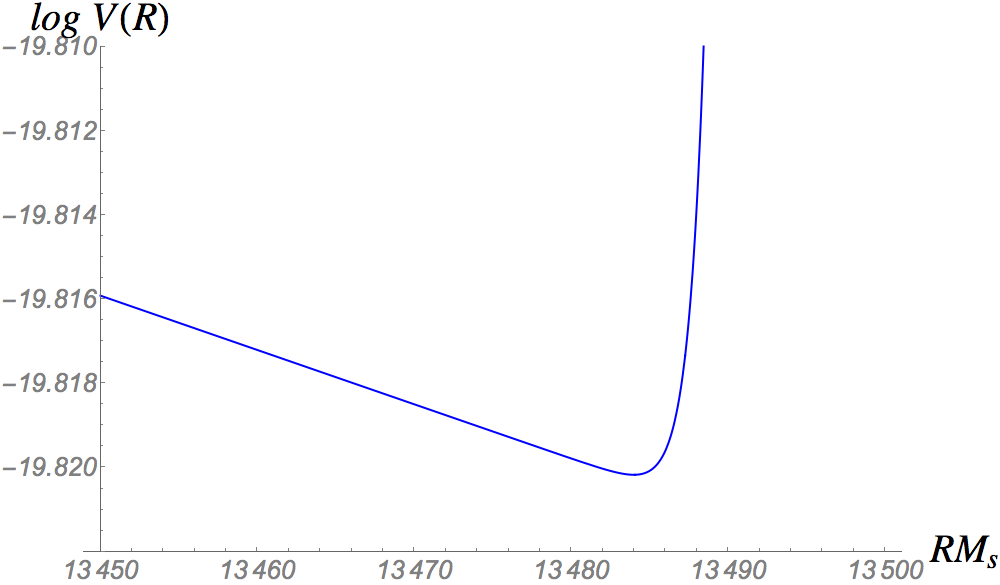}
\label{fig:fig1}
\caption{The 5D potential for $\alpha=\alpha_{D}=0.1$, $N_{f}^{0}-N_{b}^{0}=10$,
$g_{s}=3\times10^{-2}$, $C=b=-13$, $N_{c}=6$. The approximation in eq.(\ref{eq:mini}) gives the minimum at
$R_{min}=1.35\times10^{5}\ell_{s}$. As described in the text, the dynamical scale $\Lambda_e \approx 0.09 /R$ is significantly less
than the KK scale at the minimum.}
\end{figure}

As promised the minimum is automatically balanced to appear
at the correct values of $R_{min}$. An example of the potential
is shown in fig.~\ref{fig:fig1} for sample values. It is essentially a $1/R^{4}$
runaway to large radius until the ISS contribution takes over where
the SQCD gauge coupling is starting to become strong. 
The minimum is de Sitter, and
of order $10^{-3}M_{KK}^{4}$. Clearly for
consistency one would then require some additional $R$-independent
and negative contribution to bring the final cosmological constant
close to zero. 

Note that going along implicitly with need to protect the ISS mechanism from the supersymmetry breaking
of the SS mechanism, is of course the converse assumption that the effects of strong coupling in the 
SCQD sector do not disrupt the original calculation of the Casimir energy. This assumption is credible because 
the latter is dominated by the tower of KK states
with masses between $M_{KK}$ and $M_s$, and above physics occurring at the scale $\Lambda_e $ provided that the 
$\Lambda_e < 1/R_{min}$ constraint is satisfied. This condition can be relaxed in various cases and under various assumptions which 
will be made more precise 
when we come to study the string embedding in later sections. 

The form of the potential for $R\gg R_{min}$ is not well determined.
In these regions the dynamical scale is larger than the KK scale (i.e.
$R\Lambda_{e}(R)\gg 1$) so the sufficient condition in eq.(\ref{eq:hierarchy}) is violated. Most probably this implies that the 
potential turns over at some point, and the minimum
at $R_{min}$ is metastable in the $R$ direction as well, with larger
values simply reverting to runaway behaviour. It is not clear
how the large radius limit of such theories lifts to the decompactified
6D theory; most likely it is related to the 4D IR free magnetic dual
of the ISS theory, rather than the original electric SQCD theory. 

If one makes the conservative assumption that the minimum derived
above is indeed only metastable, it is important to consider what the tunnelling
rate would be to continued runaway along $R$, in order to confirm that it is sufficiently
small. An estimate requires the normalization of the modulus corresponding
to $R$. In flat space compactifications derived from string
theory the K\"ahler potential is given by $K\sim-\log V$ where $V$
is the overall compactification volume. In the present case one can
identify $V\sim i\left(T_{R}-\bar{T}_{R}\right)$ with $T_{R}$ being
a holomorphic modulus whose imaginary part gives $R$. This would
give kinetic terms for $R$ of the form $\mathcal{L}\supset\frac{|\partial T_{R}|^{2}}{R^{2}}$
so the canonically normalised field is $\phi_{R}=T_{R}/R_{min}$.
The tunnelling action can then be approximated in the thick wall limit.
The advantage of this physical situation is that the height of the
barrier does not appear in the action at leading order, only its width and the difference $\Delta V$ between the 
vacuum energies of the false and true minima.
A crude estimate for the action is then \cite{Duncan:1992ai}
\begin{equation}
S_{E}\sim2\pi^{2}\frac{\left(\Delta\phi_{R}\right)^{4}}{\Delta V}\,,
\end{equation}
where $\Delta V=V_{false}-V_{true}=R_{min}^{-4}\alpha_{D}^{2}N\left(R_{min}\Lambda_{e}\right)^{2}$.
As this is a sufficient condition, let us adopt a conservative
value for $\Delta\phi_{R}$, namely the distance in field-space between
$R_{min}$ and the point where perturbativity breaks down, $\Lambda_{e}R\sim1$,
or $\frac{\Lambda_{e}R_{min}}{\Lambda_{e}R}=\mu_{0}/M_{s}=e^{-\frac{C_{}}{b_{}}\Delta RM_{s}}$.
This gives 
\begin{equation}
\Delta\phi_{R}\approx R_{min}^{-1}\frac{b_{}}{C_{}}\log\frac{M_{s}}{\mu_{0}}\,,
\end{equation}
leading to an estimate for the tunnelling action of 
\begin{equation}
S_{E}\sim\frac{2\pi^{2}}{\alpha^{2}}\left(\frac{M_{s}}{\mu_{0}}\right)^{2}\sim10^{3}/\alpha^{2}.
\end{equation}
This is well above the $S_{E}\gtrsim400$ that is required to ensure
stability on timescales of the age of the universe (see e.g. ref.~\cite{Riotto:1995am} and references therein).
Heuristically this is simply a consequence of the fact that the Casimir
energy in $V$ is a one-loop effect (given by $\rho$), so the potential is much flatter than it is broad. 

\subsection{The exponentially suppressed ({\it UV-Casimir energy}) case}

\label{uv-casi}
Models that are non-supersymmetric but nevertheless have equal numbers
of massless bosons and fermions $(N_{f}^{0}=N_{b}^{0})$ have a 
one-loop cosmological constant that is exponentially suppressed. Ref.~\cite{Abel:2015oxa}
argues that these cases are particularly interesting due to their
enhanced stability properties, and form a better basis for doing phenomenology.

The philosophy in these cases is somewhat different: the general idea is that the 
exponential suppression appearing in the vacuum energy also appears in the scale setting the Higgs mass.  
Therefore the compactification volume (and consequently the SS supersymmetry breaking scale $1/R$) needs to be only so 
large so as to be able to generate the necessary suppression, while it is possible to live with supersymmetry breaking that is much larger than the electroweak scale (the canonical situation with SS breaking). The issue for the present discussion then is how to stabilise with exponentially small 
cosmological constant and reasonable coupling, but still with moderately large volume.
  
As already mentioned, the aspect of these theories 
that will be of particular relevance is that the only modes that make  a non-vanishing contribution to the vacuum energy have string sized masses, and indeed
the leading contribution to the Schwinger integral comes from a saddle-point at the UV end, $t \sim 1$, rather than 
from the entire integral, as is the case for a generic theory. Consequently the contribution to the cosmological constant 
resulting from the Scherk-Schwarz mechanism is blind to the IR physics occurring in for example the ISS mechanism, and the two contributions are physically separated. Indeed the former cannot easily be understood within an effective field theory\footnote{Conceivably one could try to write down a supergravity theory truncated at the first string excitation level.}.

As already mentioned, to emphasise the distinction these theories will be 
said (using the terminology in its broad sense) to have a {{\it UV-Casimir energy}}.
An additional advantage in the present context is of course that the
volumes required are much smaller than the generic case, and hence
the decompactification problem is less pronounced. 

Assuming that the exponential
suppression continues beyond one-loop, such cases have to be treated
quite differently. In the present toy model, the ISS mechanism essentially governs the minimisation, and the  issue is to  ensure that the 
contribution to the cosmological constant from the initial SS mechanism is negligible. The 5D case is as follows.

First let us return to the constraints in eq.(\ref{eq:hierarchy}).
The potential takes the form
\begin{equation}
V=R^{-4}\left[\alpha^{2}\rho(N_{f}^{1}-N_{b}^{1})(RM_s)^2e^{-4\pi RM_{s}}+N_{c}\alpha_D^2(R\Lambda_{e})^{2}\right]\, ,
\end{equation}
where $N_{f}^{1}-N_{b}^{1}$ counts the fermi-bose non-degeneracy
at the {\it first excited string level}, $\alpha$ stands again
for a generic Scherk-Schwarz phase, while $\rho \ll1$ is now generically
a one-loop suppression factor. As is evident from eq.(\ref{eq:rlam}) the 
SS term dies away rapidly at large radius. The minimum occurs shortly after the second term has its 
independent minimum at 
\begin{eqnarray}
R_{min}M_{s} & = & \frac{4b_{}}{C_{}}\,,
\end{eqnarray}
so the string scale can be perhaps an order of magnitude higher than the KK scale. It is useful to define $\sigma\gtrsim0$ as 
the final ratio of dynamical to KK scale, i.e. $R_{min}\Lambda_{e}=e^{-\sigma}$, so that ultimately 
\begin{equation}
\sigma =\frac{8\pi^2}{|b| g_s^2} - 2   \,.
\end{equation}
As usual, $m_D$ must satisfy the constraints in eq.(\ref{eq:hierarchy}), so it lies below $\Lambda_e$ but is large enough that $\hat{m}_D\Lambda_e>\alpha^2/R^2$:
\begin{eqnarray}
\label{hi2}
1\gtrsim e^{-\frac{8\pi^2}{|b| g_s^2} + 2 } &\gg& \alpha_D ,\,  \alpha^{2}/\alpha_D     \, .
\end{eqnarray}
Thus for
the mechanism to work when the Casimir energy is exponentially suppressed in the 5D$\rightarrow$4D theory, the scale of supersymmetry breaking has to be at most a few orders of magnitude below the string
scale with relatively large coupling, $8\pi^2/|b|g_s^2 \sim 2$.
In addition $\alpha_D \sim  \alpha$ are required to be small. Note that if these constraints are satisfied then the SS contribution to the cosmological constant is guaranteed to be negligible at the minimum, which was the point we wished to demonstrate here. The essential advantage of a UV-Casimir energy in this  case is that the (relatively) large volume stabilisation governed by non-perturbative long-range physics has not fed-back into it, so one has the sort of modularity normally associated with brane configurations.

 As an example, taking $|C|=10$, $|b|=30$, $g_s=1/\sqrt{2}$, one requires $\alpha,\alpha_D\lesssim R\Lambda_e\approx 1/25$.
We will later see how to accommodate $\Lambda_e > M_{KK}$ in the 
 6D$\rightarrow$4D version; this removes the upper bound in eq.(\ref{hi2}) allowing maximal SS phases $\alpha_D,\alpha\sim 1$.
 
\section{String/supergravity embedding}

\label{seciii}

Let us now collect the components for a more complete implementation within a string compactification,
focussing on a theory compactified to $\mathcal{N}=1$ in 6D, and
then further compactified on an orbifold of $\mathbb{T}_{2}$ down
to $D=4$. 

The discussion begins with a summary of the effective spontaneously
broken supergravity theory and then compares the spectrum to that
of the Scherk-Schwarzed string theory (using the framework of ref.~\cite{Abel:2015oxa}).
The extension to 6D introduces modular symmetries that persist (as a congruence subgroup) in the Scherk-Schwarzed theory.
It is shown that both the spectrum and the Casimir energy preserve these symmetries. Their great advantage is that 
they can be used to follow soft-terms in the spontaneously broken supergravity
theory (taking over the role of the $R$-symmetry 
in global SQCD \cite{Luty:1999qc,Abel:2011wv}).

This allows us to consider the theory as a whole, without
having to separate supersymmetry breaking scales with artificially
small SS twists as was done in the previous section. In fact we will ultimately 
find that the SS-induced soft-terms act to stabilise the 
minimum so that we do not have to rely on the one-loop metastability of ISS.  

\subsection{Spectrum and congruence subgroups in the effective supergravity theory}

First let us establish how the
Scherk-Schwarz mechanism in a direct string implementation such as that in ref.~\cite{Abel:2015oxa}
maps to the effective supergravity theory. As mentioned, the SS stage
of compactification is on an orbifolded $\mathbb{T}_{2}$ torus, which in the absence of 
Wilson lines can be described generally by the metric
\begin{equation}
G_{ij}=\frac{T_{2}}{U_{2}}\left(\begin{array}{cc}
1 & U_{1}\\
U_{1} & |U|^{2}
\end{array}\right)\,\,\,;\,\,\,\,G^{ij}=\frac{1}{T_{2}U_{2}}\left(\begin{array}{cc}
|U|^{2} & -U_{1}\\
-U_{1} & 1
\end{array}\right)\,,
\end{equation}
where in order to conform with most of the phenomenology oriented
SUGRA literature the convention is 
\begin{eqnarray}
iU & = & U_{1}+iU_{2}\nonumber \\
iT & = & T_{1}+iT_{2}\,.
\end{eqnarray}
For reference, untilted tori have $U_{1}=0$, $T_{2}=R_{1}R_{2}$,
$U_{2}=R_{2}/R_{1}$ where $R_{i}$ is the radius along direction
$i$, and it will be assumed throughout that $R_{2}>R_{1}$. The $U_{1}$
modulus encapsulates the tilt angle (i.e. $U_{1}=R_{2}\cos\theta/R_{1}$,
$U_{2}=R_{2}\sin\theta/R_{1}$) and $T_{2}=R_{1}R_{2}\sin\theta$ is the volume.
The nett effect on the spectrum of the Scherk-Schwarz action can be
determined on the string theory side from the shift in the internal
momenta, which can in turn be read off the partition function. The latter contains
a factor 
\begin{equation}
\mathcal{Z}_{d,d}(G,B)=\frac{1}{|\eta(\tau)|^{2d}}\sum_{\mathbf{n},\mathbf{m}}q^{\alpha'\mathbf{p_{L}^{2}/2}}\bar{q}^{\alpha'\mathbf{p_{R}^{2}/2}},\label{eq:origibos}
\end{equation}
coming from the compactified toroidal directions. The momenta depend
on the KK numbers $m_{1,2}$ and winding numbers $n^{1,2}$ of the
$\mathbb{T}_{2}$ as 
\begin{eqnarray}
\mathbf{p}_{L}^{2} & = & p_{L_{i}}G^{ij}p_{L_{j}}\nonumber \\
p_{L_{j}} & = & \frac{1}{\sqrt{2\alpha'}}\left(m_{j}+(B_{jk}+G_{jk})n^{k}\right)\,,
\end{eqnarray}
and 
\begin{eqnarray}
\mathbf{p}_{R}^{2} & = & p_{R_{i}}G^{ij}p_{R_{j}}\nonumber \\
p_{Rj} & = & \frac{1}{\sqrt{2\alpha'}}\left(m_{j}+(B_{jk}-G_{jk})n^{k}\right)\,,
\end{eqnarray}
where the notation
throughout is as in ref.~\cite{Abel:2015oxa}. The Scherk-Schwarz action causes a discrete Lorentz rotation and boost
involving the KK and winding numbers and the charge/momentum lattice,
$\mathbf{Q}$, of the form 
\begin{eqnarray}
\mathbf{Q} & \rightarrow & \mathbf{Q}-n^{i}\mathbf{e}_{i}\nonumber \\
m_{i} & \rightarrow & m_{i}+\mathbf{Q}\cdot\mathbf{e}_{i}-\frac{1}{2}\mathbf{e}_{i}\cdot\mathbf{e}_{j}n^{j}\nonumber \\
B_{jk}\pm G_{jk} & \rightarrow & B_{jk}\pm G_{jk}-\frac{1}{4}\mathbf{e}_{j}\cdot\mathbf{e}_{k}\,\,\,,\label{eq:Wilson}
\end{eqnarray}
where $\mathbf{e}_{i=1,2}$ are vectors containing the Scherk-Schwarz
action on the $R$-charges and possibly also gauge charges, and the
dot product refers to the Lorentzian charge lattice. The vectors $\mathbf{e}_{i}$
contain the phases $\alpha_{F,B}$, although one should note they
must leave the world sheet supercurrent and charge lattice invariant, and 
have to leave a consistent orbifold projection.
It is for these reasons that $\alpha_{F,B}$ are constrained to be discrete. 

Specialising to the maximal twist case, the spontaneous supersymmetry
breaking arises from half integer values of the $\mathbf{Q}\cdot\mathbf{e}_{i}$
shift in the KK numbers. Consider the gravitinos; adding left and
right moving contributions, the modes $m_{i}$ and $\mathbf{Q}\cdot\mathbf{e}_{i}$
marry with the modes $-m$ and $-\mathbf{Q}\cdot\mathbf{e}_{i}$,
so that
\begin{eqnarray}
\left(m_{3/2}^{(m_{1}m_{2})}\right)^{2} & = & \frac{1}{\alpha'}\hat{m}_{i}G^{ij}\hat{m}_{j}=\frac{1}{4\alpha'}\sum_{ij}G^{ij}\nonumber \\
 & = & \frac{1}{\alpha'}\frac{1}{T_{2}U_{2}}\left|(m_{1}-\frac{1}{2})-(m_{2}-\frac{1}{2})iU\right|^{2}\,.\label{eq:gengrav}
\end{eqnarray}
Clearly supersymmetry is restored for all $U_{1}=\frac{2\ell_{1}-1}{2\ell_{2}-1}$
in the limit $U_{2}\rightarrow0$ for integer $\ell_{1,2}$. This
limit can be achieved by decompactifying with constant ratio of radii, with the tilt
angle going to zero (slower than $1/R_1R_2$ in order for $T_2$ to go to large volume). An identical mass-shift is induced in the gauginos.
From this we can identify the effective KK scale near a supersymmetric
point as $M_{KK}^{2}=U_{2}/T_{2}=1/R_{1}^{2}$. (Where necessary factors
of $\alpha'$ are absorbed into the modulus $T$ to give it dimensions
of length squared.)

Continuous Wilson lines shift the KK and winding numbers along with
the internal charges in a similar fashion and these can be related
to matter/Higgs fields: the shift induced by the pair of continuous
real Wilson lines $\mathbf{A}_{1}$, $\mathbf{A}_{2}$, can be written
\begin{eqnarray}
\mathbf{Q} & \rightarrow & \mathbf{Q}+n^{i}\mathbf{A}_{i}\nonumber \\
m_{i} & \rightarrow & m_{i}-\mathbf{Q}\cdot\mathbf{A}_{i}\nonumber \\
B_{jk}\pm G_{jk} & \rightarrow & B_{jk}\pm G_{jk}-\frac{1}{4}\mathbf{A}_{j}\cdot\mathbf{A}_{k}\,\,\,.\label{eq:Wilson-1}
\end{eqnarray}
The real shift vectors $\mathbf{A}_{i}$ can be related to a pair
of complex fields in the effective supergravity theory, denoted $\phi$,
$\phi'$. To get to this basis, first define complex Wilson lines,
\begin{equation}
\mathbf{Z}=iU\mathbf{A}_{1}-\mathbf{A}_{2}\,,
\end{equation}
and then 
\begin{eqnarray}
i\phi & = & \frac{1}{2}(Z^{1}-iZ^{2})\nonumber \\
i\phi' & = & \frac{1}{2}(Z^{1}+iZ^{2})\,.
\end{eqnarray}
The upper indices refer to basis vectors for the charge lattice. Defining
$2P=\phi+\bar{\phi'}$, useful combinations are (in our conventions)
\begin{eqnarray}
P\bar{P} & = & \sum_{a}\Im(Z^{a})^{2}=\left(\mathbf{A}_{1}\cdot\mathbf{A}_{1}\right)\frac{U_{2}^{2}}{4}\nonumber \\
T_{2} & = & \sqrt{G}+P\bar{P}/U_{2}\nonumber \\
U & = & \frac{1}{G_{11}}\left(\sqrt{G}-iG_{12}\right)\,,
\label{eq:newdet}
\end{eqnarray}
with the $T_{2}$ redefinition matching the shift in (\ref{eq:Wilson-1}).
Going from $\mathbf{Z}$ to $\phi$, $\phi'$ amounts to a change
of basis for $\mathbf{Q}$. For example the current superfield for
a $U(1)$ current (under which $\phi$ and $\phi'$ must have opposite
charges) is given by, $J=|\phi|^{2}-|\phi'|^{2}=\frac{i}{2}\left(Z^{1}\bar{Z}^{2}-Z^{2}\bar{Z}^{1}\right)$,
so its generator acts as $SO(2)$ on the $Z^{a}$ indices. The K\"ahler potential depends on the volume $\sqrt{G}$ as 
\begin{equation}
K=-\log Y-\log4(T_{2}U_{2}-P\bar{P})\,,\label{eq:kahler}
\end{equation}
where 
$2P=\phi+\bar{\phi'}$, and where the dilaton combination generally includes a term from the (heterotic)
Green-Schwarz mechanism, 
\begin{equation}
Y=S+\bar{S}-\delta_{GS}\log4(T_{2}U_{2}-P\bar{P})\,.
\end{equation}

So far the picture is just that of the standard $\mathcal{N}=1$ theories,
but now we deform the theory with a superpotential
that successfully reproduces the SSSB observed in the string spectrum.
As we saw on the string side in eq.(\ref{eq:gengrav}), near $iU=1$
the lightest spin 3/2 state is the zero-KK mode gravitino whose physical
mass is 
\begin{equation}
\label{m32}
m_{3/2}^{2}=\frac{1}{4}\frac{1}{S_{2}T_{2}U_{2}}|1-iU|^{2}.
\end{equation}
The relation between the Planck scale and string scale is
\begin{equation}
M_{P}^{2}=g_{s}^{-2}\alpha'^{-1}\,,
\end{equation}
which suggests that a superpotential in the spontaneously broken theory
that produces the correct spectrum is
\begin{equation}
W_{SS}=\sqrt{2}(1-iU)\,.\label{eq:shtom}
\end{equation}

It can be verified that near $U_{1}=1$, the rest of the low-lying
tree-level string spectrum is successfully generated by this supergravity
theory. Explicitly, in the string spectrum the {\it tree-level}
gaugino masses are degenerate with the gravitino: using standard notation,
the supercovariant derivative is $D_{i}W=W_{i}+WK_{i}$, and the gauge
kinetic function is $f_{tree}=S$, leading to 
\begin{equation}
m_{\lambda}=\left|\frac{m_{3/2}}{2}Re(f_{tree})^{-1}K^{\bar{i}j}\partial_{i}f_{tree}\frac{D_{\bar{j}}\bar{W}}{\bar{W}}\right|=m_{3/2}\,.\label{eq:gagmass}
\end{equation}
At one-loop the masses would not be equal in either the field theory
or the string theory due to gauge mediation effects, but we shall
see below that the above relation does not suffer large volume corrections.

Continuing the comparison of the spectra, after spontaneous superymmetry
breaking {\it all} the untwisted scalars in the NS-NS sector should
remain massless at tree-level, while their fermion superpartners
pick up a mass equal to that of the gravitino. The corresponding superfields,
$\phi$ and $\phi'$, achieve this by appearing to conspire in the
K\"ahler potential as 
\begin{eqnarray}
K & \supset & -\log\left(4T_{2}U_{2}-|\phi+\bar{\phi'}|^{2}\right)\nonumber \\
 & = & -\log4T_{2}U_{2}+\frac{1}{4T_{2}U_{2}}(|\phi|^{2}+|\phi'|^{2}+\phi\phi'+\bar{\phi}\bar{\phi'})+\ldots
\end{eqnarray}
The tree-level fermion mass terms in the effective theory (which is
a ``$\mu$-term'' if one is thinking of $\phi$,$\phi'$ as Higgses),
are then given by

\begin{equation}
\mu_{\phi}=m_{3/2}Z_{\phi}^{-\frac{1}{2}}Z_{\phi'}^{-\frac{1}{2}}\left(\frac{W_{ij}}{W}+K_{ij}-\Gamma_{ij}^{k}\frac{D_{k}W}{W}\right)\,,
\end{equation}
where 
\begin{eqnarray}
\Gamma_{ij}^{k} & = & K^{k\bar{k}}\partial_{\bar{k}}K_{ij}\nonumber \\
Z_{\phi}^{-1}=Z_{\phi'}^{-1} & = & 4T_{2}U_{2}=1/K_{\phi\phi'}\,.
\end{eqnarray}
Inserting the supersymmetry breaking superpotential in eq.(\ref{eq:shtom})
gives
\begin{equation}
\mu_{\phi}=m_{3/2}\left(4T_{2}U_{2}\frac{W_{ij}}{W}-\frac{\bar{W}}{W}\right)\,.
\end{equation}
In the absence of any explicit $W_{ij}$ mass terms in the original
superpotential, this automatically has the same magnitude as the gaugino
and gravitino masses in accord with the Scherk-Schwarzed string theory
spectrum. It is straightforward to show that $S,U,T,\phi$ and $\phi'$
fit into a larger ``no-scale'' supergravity structure that leaves
{\it all} the scalars massless at tree-level (modulo possible variations
in the splittings of the matter fields that may arise if $\mathbf{e}$
is also embedded into the gauge groups: in the effective theory this
would correspond to turning on scalar ``Higgs'' VEVs). The conspiring
dimensionful terms correspond to mass-squareds and Dirac masses of
magnitude $m_{3/2}$ for the {canonically normalized states.} 

\begin{figure}
{\mbox{\qquad\qquad\qquad\qquad\qquad\qquad\qquad\qquad\qquad\qquad\qquad\qquad\qquad\qquad}{\large $\lfloor iU$}}
 \vspace{0.1cm}\\
\noindent \begin{centering}
\includegraphics[bb=180bp 0bp 950bp 490bp,clip,scale=0.35]{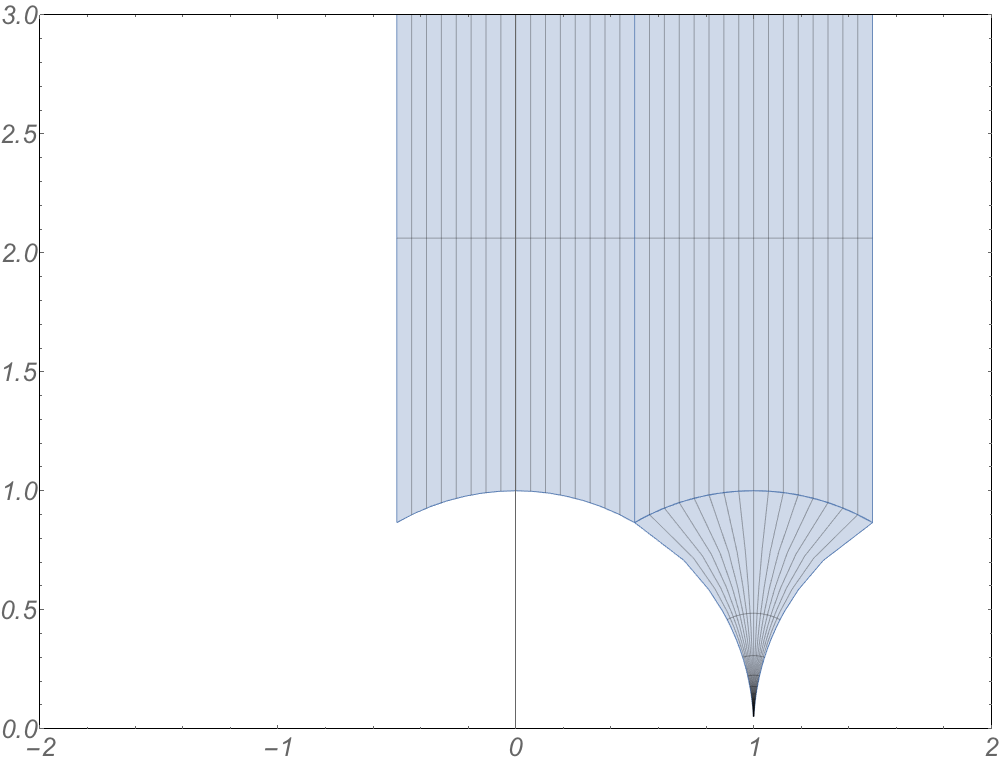}
\par\end{centering}
\noindent \centering{}\caption{\label{fig:E3-1}The fundamental $U$-modulus domain for a maximally
twisted Scherk-Schwarz theory has a supersymmetric cusp at $iU=1$. }
\end{figure}

The original ${\cal N}=1$ theory has well-known modular symmetries: for completeness the standard $SL(2,\mathbb{Z})_T$ 
and $SL(2,\mathbb{Z})_U$ symmetries of 
the  supersymmetric theory are included in the Appendix. What remains of them after applying the
Scherk-Schwarz mechanism? Due to the spontaneous nature of the breaking it is clear that the 
K\"ahler potential should still respect the full symmetry, as it indeed does, and that the new SSSB superpotential should be the only source of its breaking. 
To see its effect on the modular symmetries consider the spectrum: 
according to eq.(\ref{eq:gengrav}) the
zero-mode KK gravitino need not be the lightest state, depending on
the value of $U_{1}$. If $U_{1}=\frac{2\ell_{1}-1}{2\ell_{2}-1}$
then the lightest gravitino is instead the $\ell_{1},\ell_{2}$ KK
mode for all $U_{2}\lesssim1/(2\ell_{2}+1)$, and the superpotential
in the effective theory would actually be $W_{SS}=\sqrt{2}((2\ell_{1}-1)-(2\ell_{2}-1)iU)$
near this point. The fact that one has to specify which mode plays
the role of the gravitino in the effective theory is of course just a symptom
of the deficiency of the 4D supergravity approximation, which 
cannot describe the supersymmetry breaking over the whole $U$ moduli-space.
Indeed the explicit breaking of modular symmetry in the superpotential
just amounts to a choice of gauge: because of the original discrete
symmetry, there are infinitely many equivalent spontaneously broken
theories that one could write down for the effective supergravity
theory related by a subgroup of the $SL(2,\mathbb{Z})_{U}$ transformations.
This is evident from the fact that under transformations of the form
\begin{equation}
\frac{1}{4}\frac{|1-iU|^{2}}{S_{2}T_{2}U_{2}}\equiv\frac{1}{4}\frac{|(d-b)-(a-c)iU|^{2}}{S_{2}T_{2}U_{2}}\,\,\,;\,\,a,b,c,d\in\mathbb{Z},\,\,ad-bc=1,\,\,a-c=b-d=1\,\mbox{mod}(1)\,,
\end{equation}
the gravitino spectrum is invariant. In fact the entire theory
is invariant only under the smaller congruence subgroup defined by
$a,d=1$ mod~(1) and $b,c=0$ mod~(1), similar to ref.~\cite{Angelantonj:2015nfa},
which will be referred to as $\Gamma_{\vartheta}(2)$. Under such transformations,
any $U$ in a maximally twisted Scherk-Schwarz theory can be mapped
to the fundamental domain shown in fig.~\ref{fig:E3-1}. In addition
to the cusp at infinity, there is a single representative supersymmetric
cusp at $iU=1$. For non-maximal Scherk-Schwarz twists, the fundamental
domain will contain more cusps, and there will be several genuinely
distinct supersymmetric vacua (consult ref.~\cite{Angelantonj:2015nfa}
for details). Naturally the Casimir energy, when we come to calculate it, must
respect this symmetry.

We will also need an understanding of
the one-loop gauge thresholds. Their volume dependence (neglecting
the effects of extra charged massless states) can be written \cite{Angelantonj:2015nfa}
\begin{equation}
\Delta_{}=-C_{}\log\left(T_{2}U_{2}|\eta(iT)|^{4}|\eta(iU)|^{4}\right)+(C_{}-b_{})\log\left(T_{2}U_{2}|\vartheta_{4}(iT)|^{4}|\vartheta_{2}(iU)|^{4}\right)\,,\label{eq:thresh}
\end{equation}
where $b_{}=16\pi^{2}\beta$ is the beta function coefficients
for the entire massless theory, $C_{}=16\pi^{2}\beta_{\mathcal{N}=2}$
is the $N=2$ coefficient, and $\eta$ are the usual Dedekind eta
functions. The modular functions in this expression are also invariant
under $\Gamma_{\vartheta}(2)$ transformations; denoting $SL(2,\mathbb{Z})_{U}$
operations by $S_{U}\equiv iU\rightarrow-1/iU$ and $T_{U}\equiv iU\rightarrow iU+1$,
we have 
\begin{eqnarray}
T_{U}:U_{2}|\vartheta_{2}(iU)|^{4} & \longrightarrow & U_{2}|\vartheta_{2}(iU)|^{4}\\
S_{U}:U_{2}|\vartheta_{2,4}(iU)|^{4} & \longrightarrow & U_{2}|\vartheta_{4,2}(iU)|^{4}\,.
\end{eqnarray}
Therefore $\Delta_{}$ is invariant under any number of $T_{U}$
moves, but only an even number of $S_{U}$ moves, in accord with the
congruence condition. 

Following now the standard route (see for example refs.~\cite{Kaplunovsky:1987rp,Kaplunovsky:1995jw,Lalak:1999bk,Parameswaran:2010ec})
this allows us to identify the holomorphic gauge kinetic function
of the SQCD as (taking a Kac-Moody level $k=1$ for the gauge group),
\begin{equation}
f_{}=S-\frac{C_{}}{8\pi^{2}}\log\eta(iT)^{2}\eta(iU)^{2}+\frac{C_{}-b_{}}{8\pi^{2}}\log\left(\vartheta_{4}(iT)^{2}\vartheta_{2}(iU)^{2}\right)\,,
\end{equation}
with the gauge coupling being given by 
\begin{equation}
\frac{2}{g^{2}}=Y=2\Re(f_{})-\frac{b_{}}{8\pi^{2}}\log(\mu^{2})-\left(\frac{b_{}}{8\pi^{2}}+\delta_{GS}\right)\log(4T_{2}U_{2})\,.\label{eq:foo1}
\end{equation}
Note that due to the additional universal terms it is the $\mathcal{N}=1$ beta function appearing here
(i.e. $b_{}=-3N+F$ in $SU(N)$ gauge theories with $\mathcal{N}=1$
SQCD and $F$ flavours), and
not $C_{}$. 

The holomorphic dynamical scale $\Lambda_{hol}$ can be defined as
\begin{equation}
\Lambda_{hol}=\exp\left(-\frac{8\pi^{2}}{|b_{}|}f_{}\right)\,,\label{eq:lhol}
\end{equation}
and the modular weight of $\Lambda_{hol}$ is given by 
\begin{equation}
n_{\Lambda}=8\pi^{2}\frac{b_{}/8\pi^{2}+\delta_{GS}}{|b_{}|}\, .\label{eq:foo2}
\end{equation}
The gauge coupling can then be written more succinctly as 
\begin{equation}
\frac{1}{g^{2}(\mu)}=-\frac{b_{}}{8\pi^{2}}\log\left[\frac{\mu}{|\Lambda_{hol}|(4T_{2}U_{2})^{n_{\Lambda}/2}}\right]\,.\label{eq:foo1-1}
\end{equation}
It will often be useful to leave $n_{\Lambda}$ implicit, as it is
essentially just whatever combination of terms appears in
eq.(\ref{eq:foo1}). However it can be calculated directly \cite{Kaplunovsky:1995jw};
specialising to $SU(N)$ gauge theories with $\mathcal{N}=1$ SQCD
and $F$ flavours of quark and anti-quark, it is 
\begin{equation}
|b_{}|n_{\Lambda}=2Fn_{Q}+F-N\,.\label{eq:nlrel}
\end{equation}
We will see that this equation provides an important consistency condition 
for the implementation of the ISS mechanism, because it can be derived independently from 
the matching conditions for the Seiberg duals.
Note that it will be assumed for simplicity throughout that the $SL(2,\mathbb{Z})_{U}$
and $SL(2,\mathbb{Z})_{T}$ weights are degenerate for every field.

To complete this part of the discussion, one can obtain an asymptotic approximation for the gauge
threshold correction at large volume and in the supersymmetric limit around the representative 
 cusp at $iU=1$ (which obviously breaks the modular symmetry).
In the vicinity of the cusp, since $\lim_{iU\rightarrow1}\eta(iU)=0$,
it is often convenient to use $SL(2,\mathbb{Z})_{U}$ modular redefinitions
to the cusp at infinity, that is $i\tilde{U}=-1/(iU-1)\approx i/U_{2}$,
with $i\tilde{U}\rightarrow i\infty$ in the supersymmetric limit:
the standard expansion $\vartheta_{4}(i\tilde{U})\rightarrow1-2e^{-\pi\tilde{U}}+\ldots$
then gives, 
\begin{eqnarray}
\Delta_{} & = & -C_{}\log\left(4T_{2}\tilde{U}_{2}|\eta(iT)|^{4}|\eta(i\tilde{U})|^{4}\right)+(C_{}-b_{})\log\left(4T_{2}\tilde{U}_{2}|\vartheta_{4}(iT)|^{4}|\vartheta_{4}(i\tilde{U})|^{4}\right)\,,\nonumber \\
 & = & \frac{\pi}{3}C_{}\left(T_{2}+\tilde{U}_{2}\right)-b_{}\log\left(4T_{2}\tilde{U}_{2}\right)+\mathcal{O}(e^{-\pi\tilde{U}_{2}},e^{-\pi T_{2}})\,.\label{eq:largevolD}
\end{eqnarray}
As in the 5D case,
the second term subtracts from $16\pi^{2}/g(\mu)^{2}$ the logarithmic
running between the lightest KK-mode $M_{KK}=1/\sqrt{4T_{2}\tilde{U}_{2}}$
and the string scale, whilst the first term replaces it with a power-law
threshold. Under our assumption that $C_{}/b_{}>0$, it is clear
that one is prevented from going continuously to the boundary of moduli-space by
the appearance of strong coupling in the QCD theory where $\frac{\pi}{3}C_{}\left(T_{2}+\tilde{U}_{2}\right)\sim16\pi^{2}$, and this is precisely 
the region in which the minimum is expected to appear.

Returning to the appearance of the large volume dependence in the
one-loop gaugino mass, retaining only the pieces $f_{}\approx S+\frac{C_{}}{8\pi^{2}}\frac{\pi}{6}\left(T+\tilde{U}\right)$,
eq.(\ref{eq:gagmass}) and a little work shows that the relation $m_{\lambda}=m_{3/2}$
holds at one-loop up to logarithmic corrections, as promised. 

\subsection{Calculation of Casimir energy}

\label{sec:casi}

Next let us determine the cosmological constant for the general $6D\rightarrow4D$
case, essentially repeating the computation
of ref.~\cite{Abel:2015oxa} in the full string theory, but now  retaining
the full $T,U$ dependence. In particular it will be possible to check that the result respects
the $\Gamma_{\vartheta}(2)$ symmetry of the congruence subgroup described
above. 

The required expression is 
\begin{equation}
\Lambda^{(4)}(T,U)=-\frac{1}{2}\int_{\mathcal{F}}\frac{d^{2}\tau}{\tau_{2}^{2}}\mathcal{Z}(\tau)\,.\label{eq:cosmo}
\end{equation}
Using the result in eq.(\ref{eq:Wilson}), the partition function
can be approximated at large volume ($T_{2}\gg1$) by neglecting the
winding modes and Poisson resumming the KK modes of eq.(\ref{eq:origibos}),
giving 

\begin{equation}
\mathcal{Z}_{\mathbf{0},\bfell}=\frac{\mathcal{M}^{2}}{\tau_{2}|\eta|^{4}}\sqrt{\det G}e^{-\frac{\pi}{\tau_{2}}\ell^{i}G_{ij}\ell^{j}}.
\end{equation}
The main simplifying approximation we are making is to neglect the
non-zero winding mode contributions (i.e. $\mathcal{Z}_{n\neq\mathbf{0},\bfell}$)
because they are suppressed by exponential factors when the volume
is large. Indeed the largest possible terms with non-zero winding
would come from otherwise massless modes with $n_{i}=1$, and would
be proportional to $\sim e^{-\pi T_{2}}/\pi T_{2}$. This should be
compared to the leading $n_{i}=0$ contributions which as in ref.~\cite{Abel:2015oxa}
have a milder exponential suppression factor of $e^{-2\pi\sqrt{T_{2}}}$.
The $n_{i}=0,\,\,\sum_{i}\ell_{i}=${\it even} contributions remain
supersymmetric regardless of the presence or otherwise of Wilson lines
(assuming the latter do not themselves break supersymmetry), and therefore
we need only consider {\it $\ell_{1}+\ell_{2}=$odd}. In addition
one can ignore the various twisted sectors of the orbifold which, being
independent of the moduli, are supersymmetric and cannot contribute
to $\Lambda$. As a further approximation one may at large volume
neglect the non-level matched terms which allows one to express the
result entirely in terms of physical states; the leading contributions
being neglected in this latter approximation are from the proto-graviton
state described in ref.~\cite{Abel:2015oxa}, and are of order $\sim T_{2}e^{-\pi T_{2}}$.
In making these approximations one obviously at this point has to
abandon the full $SL(2,\mathbb{Z})_{T}$ modular structure of $\Lambda(T,U)$,
but the $\Gamma_{\vartheta}(2)$ $U$-symmetry should remain. We are henceforth
obliged to always work at large $T_{2}$ (which just affirms the preamble
concerning the importance of interpolation).

\begin{figure}
\noindent \begin{centering}
a)\includegraphics[bb=0bp 0bp 1000bp 1000bp,clip,scale=0.2]{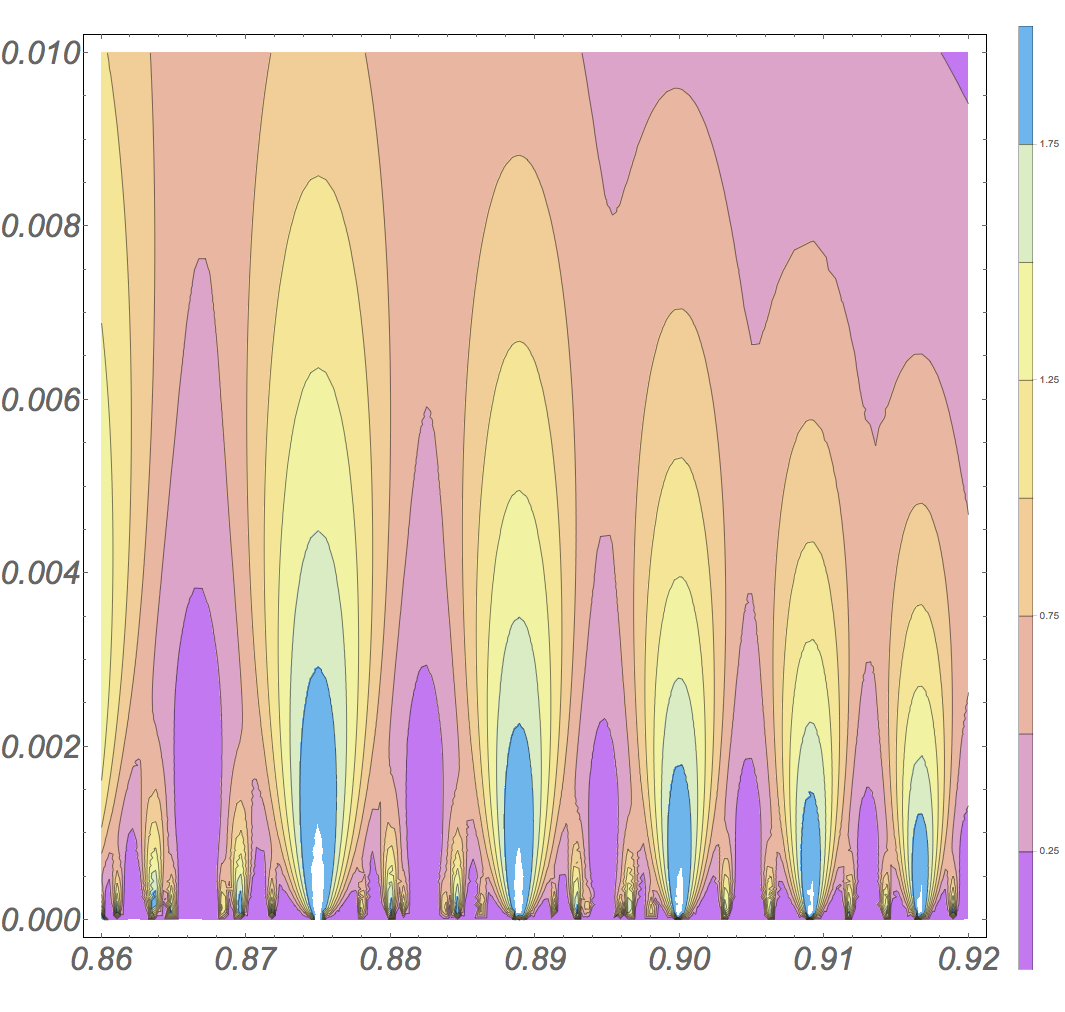}~~b)\includegraphics[bb=0bp 0bp 1000bp 1000bp,clip,scale=0.2]{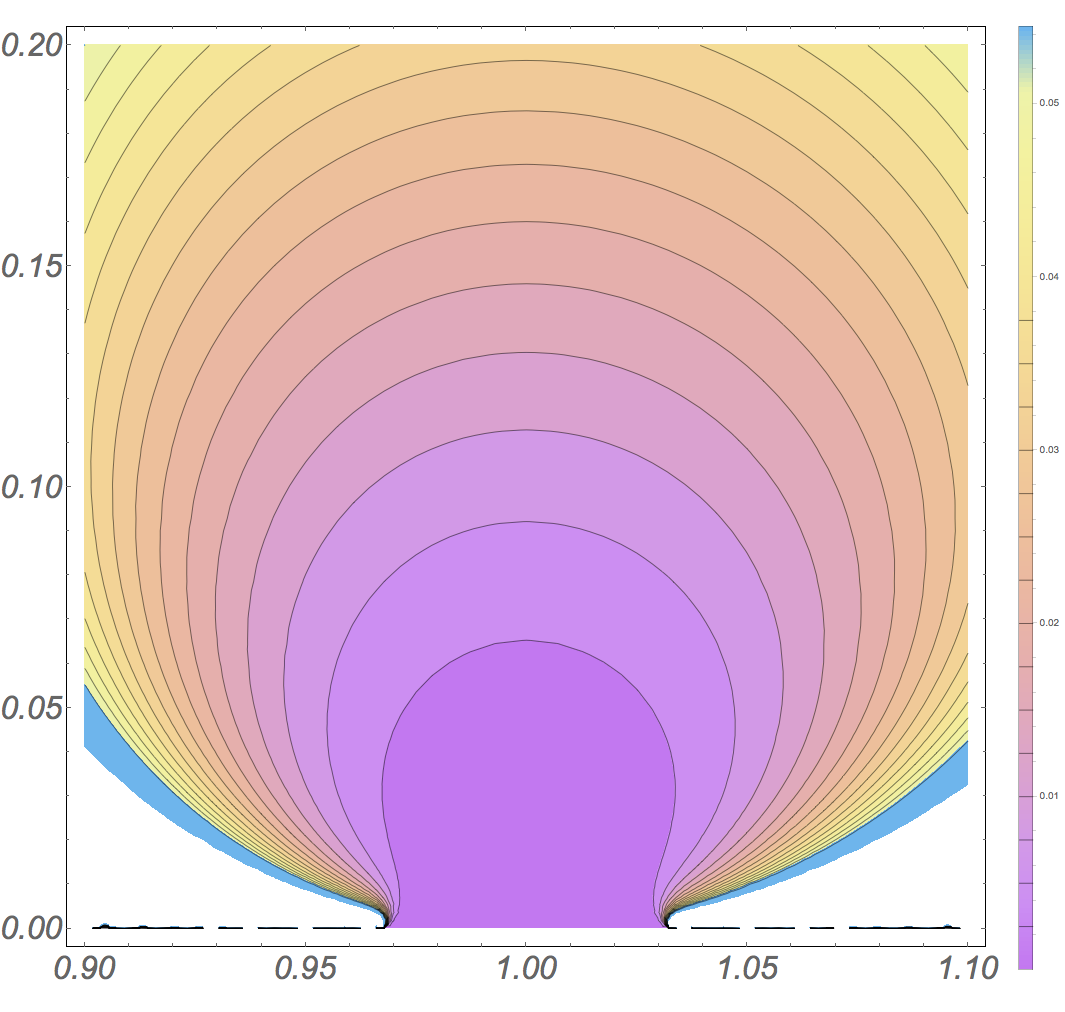}\\
~\\
$E_{3}(e^{i\theta})$
\par\end{centering}
\noindent \centering{}c)\includegraphics[bb=0bp 0bp 1000bp 600bp,clip,scale=0.3]{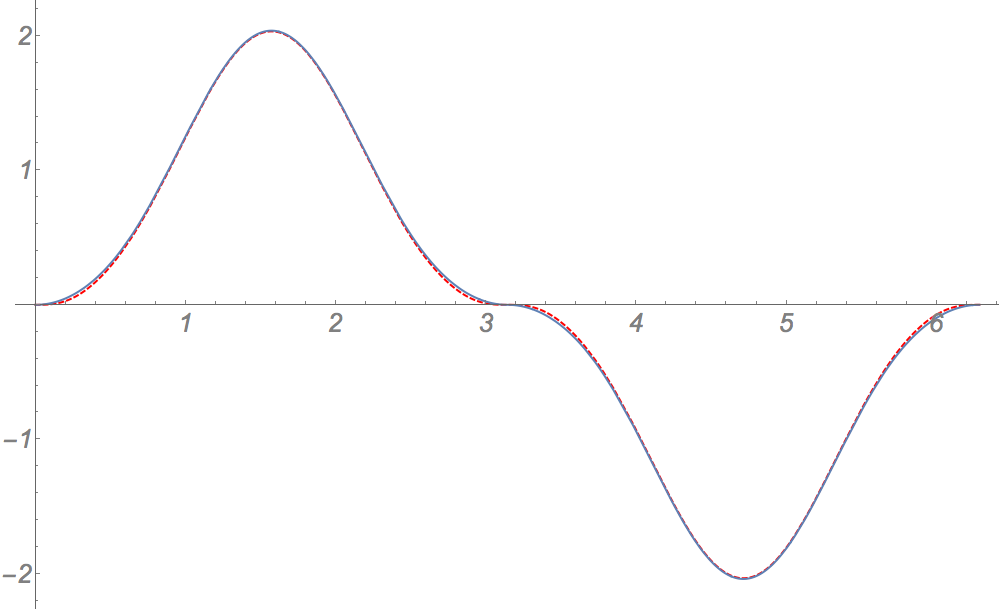}\\
\caption{\label{fig:E3}The Casimir energy $E_{3}(iU)$. In a) we see the self-similarity
near the critical line, with the bottom of each valley corresponding
to $U_{1}=(2\ell_{1}+1)/(2\ell_{2}+1)$ for integer $\ell_{1}$, $\ell_{2}$,
and a different gravitino. Fig. b) shows the vacuum energy around
$iU=1$ as a function of $\rho,\theta$ where $iU=\rho e^{i\theta}$,
and fig. c) shows it along the unit circle $iU=e^{i\theta}$. The
dashed line is the approximation $E_{3}(iU)\approx2\sum_{k}\frac{U_{2}^{3}}{|2k+iU|^{6}}$. }
\end{figure}

The result is an expression for the partition function of the form
\begin{equation}
\mathcal{Z}(\tau)\approx\frac{\mathcal{M}^{2}}{\tau_{2}|\eta|^{4}}\frac{1}{\eta^{8}\bar{\eta}^{20}}\sum_{\mathbf{\bfell}}\mathcal{Z}_{\mathbf{0},\mathbf{\bfell}}\sum_{\alpha,\beta}e^{2\pi i\sum_{i}\ell_{i}\left[\mathbf{e\cdot Q}\right]}\mathcal{Z}_{internal}\left[\begin{array}{c}
\alpha\\
\beta
\end{array}\right],
\end{equation}
where $\alpha,\beta$ label the sectors along the two cycles of the
torus. Written as a sum over the physical states this reduces to 
\begin{equation}
\label{zzz}
\mathcal{Z}(\tau)\equiv\frac{T_{2}}{\tau_{2}^{2}}\sum_{\stackrel{\bfell=odd}{level=k}}(N_{b}^{(k)}-N_{f}^{(k)})e^{-\frac{\pi}{\tau_{2}}\ell^{i}G_{ij}\ell^{j}}e^{-\pi\tau_{2}\alpha'm_{k}^{2}}\, ,
\end{equation}
where $(N_{b}^{(k)}-N_{f}^{(k)})$ is the Bose-Fermi non-degeneracy
of the {\it states unshifted by the Scherk-Schwarz mechanism at level}
$k$. Inserting into eq.(\ref{eq:cosmo}) this gives a leading contribution
to the cosmological constant of 
\begin{equation}
\Lambda(T,U)=\frac{2}{\pi^{3}}\frac{1}{T_{2}^{2}}(N_{f}^{0}-N_{b}^{0})\left[\frac{1}{2}\sum_{\ell_{1}+\ell_{2}=odd}\frac{U_{2}^{3}}{|\ell_{1}+iU\ell_{2}|^{6}}\right]\,.\label{eq:master}
\end{equation}
The sum in the square brackets, which will be referred to as $E_{3}(iU)$,
is an Eisenstein series, restricted to odd $\ell_{1}+\ell_{2}=1$ mod~(1), instead of
the canonical $(\ell_{1},\ell_{2})\neq(0,0)$. One can easily see
that it indeed respects the congruence subgroup obeyed by the spectrum,
and also that it has zeros at the supersymmetric points: indeed since
$U_{1}=(2\ell_{1}+1)/(2\ell_{2}+1)$ implies $|m_{1}+U_{1}m_{2}|\geq1/(2\ell_{2}+1)\,\forall\ell_{1}+\ell_{2}\,\,  \mbox{mod}~(1)\, =\, 1\,$,
one may smoothly take the $U_{2}\rightarrow0$ limit of the sum for
precisely these values. In accord with the modular transformation
above, there is an infinite number of such ``trivial zeros'', at
all odd integer values of $U_{1}$ as well as fractions with odd numerator
and denominator, with the general structure as one approaches the
$U_{2}=0$ line becoming extremely intricate to reflect its modular
symmetry, as shown in fig.~\ref{fig:E3}. (It is not clear if anything
interesting happens at irrational values of $U_{1}$.) 

For use in the minimisation let us focus on the Casimir energy around
the representative supersymmetric cusp at $iU=1$. The potential near
$iU=1$ is shown in fig.~\ref{fig:E3}. Clearly the minimisation
will take place near $|U|=1$ and the phase of $U$ will be the dynamically
important variable. The potential along the unit circle is also shown,
along with the following approximation which can be evaluated in closed
form: 
\begin{equation}
E_{3}(iU)\approx2\sum_{k}\frac{U_{2}^{3}}{|2k+iU|^{6}}\rightarrow\frac{\pi^{6}U_{2}^{3}}{240}\,.\label{eq:e3approx}
\end{equation}

The $N_f^0=N_b^0$ case is instead dominated by the leading saddle point. 
According to eqs.(\ref{eq:cosmo}) and (\ref{zzz}) we find 
\begin{eqnarray}
\Lambda(T,U) &=& \frac{T_{2}}{2}(N_{f}^{1}-N_{b}^{1})\sum_{\ell_{1}+\ell_{2}=odd} (\ell^iG_{ij}\ell^j)^{-7/4} e^{-2\pi \sqrt{ \ell^iG_{ij}\ell^j } }\nonumber \\
&=& \frac{(N_{f}^{1}-N_{b}^{1})}{2}\,\,T^{-3/4}_{2}U^{7/4}_{2}       \sum_{\ell_{1}+\ell_{2}=odd} 
 \frac{ e^{-2\pi \sqrt{T_2/U_2} |\ell_1+iU\ell_2|} } {|\ell_1+iU\ell_2|^{7/2}}\,.\label{eq:master-exp}
\end{eqnarray}
Expanding about $iU\approx 1$ the following approximation will be useful:
\begin{equation}
\Lambda(T,U) = 2{(N_{f}^{1}-N_{b}^{1})}\,T^{-3/4}_{2}U^{7/4}_{2}   e^{-2\pi \sqrt{T_2/U_2}  }\left( 1+{\cal O}(iU-1)\right) \,.\label{eq:master-exp}
\end{equation}

\subsection{The congruence subgroup method for mapping soft-terms between Seiberg
duals}

Next we determine how the ISS mechanism is governed by the congruence subgroup. This subsection contains two
new results. First it is shown that the string relation between
the modular weights in eq.(\ref{eq:nlrel}) can be derived as the
unique solution to modular invariance in a pair of Seiberg duals, and secondly it is shown that the congruence subgroup provides a useful means of tracking
soft-terms, including the effect of gravity mediation. It is also shown that the ISS mechanism still operates,
with all masses, dynamical scales and so forth being replaced by the
corresponding physical and hence modular invariant quantities. The issue of how KK modes 
enter into the ISS mechanism will be addressed in the following subsection.

Recall that in the ISS mechanism, the original electric theory has
a Dirac mass superpotential, 
\begin{equation}
\label{mag0}
W_{el}=m_{D}Q\tilde{Q}\,,
\end{equation}
while the magnetic dual has a superpotential 
\begin{equation}
\label{mag1}
W_{mag}=\frac{[Q\tilde{Q}]q\tilde{q}}{\hat{\Lambda}}+m_{D}[Q\tilde{Q}]\,.
\end{equation}
The inverse coupling $\hat{\Lambda}$ in the superpotential  is expected
to be of order the strong coupling scale of the theory. One can determine its modular weight from the requirement that
$W_{mag}$ has weight $-1$, as does the dynamically induced superpotential
for the SQCD theory,
\begin{equation}
W_{dyn}=-\tilde{N}\left(\frac{\det_{F}\left[Q\tilde{Q}\right]}{\hat{\Lambda}^{3N-F}}\right)^{1/\tilde{N}}\, .
\end{equation}
This yields the modular weights of $Q$ and $q$ in terms of the weight
of $\hat{\Lambda}$: 
\begin{eqnarray}
n_{q} & = & n_{\hat{\Lambda}}\frac{2F-3N}{2F}-\frac{N}{2F}\nonumber \\
n_{Q} & = & n_{\hat{\Lambda}}\frac{3N-F}{2F}-\frac{F-N}{2F}\,.\label{eq:nqs}
\end{eqnarray}
{\it A nontrivial consistency check is that these expressions are
in accord with the string relations in eq.(\ref{eq:nlrel}) in both
the electric and magnetic phases. }They are also in accord with the
well known matching relation, 
\begin{equation}
\label{mat1}
\Lambda_{hol}^{-b_{}}\tilde{\Lambda}_{hol}^{-\tilde{b}_{}}\sim\hat{\Lambda}^{-F},
\end{equation}
as well as the matching of baryons, 
\begin{equation}
\label{mat2}
\left(\frac{Q}{\Lambda_{hol}}\right)^{N}\sim\left(\frac{q}{\tilde{\Lambda}_{hol}}\right)^{\tilde{N}}\,,
\end{equation}
provided that $n_{\hat{\Lambda}_{hol}}=n_{\Lambda_{hol}}=n_{\tilde{\Lambda}_{hol}}$,
where $\Lambda_{hol}$ and $\tilde{\Lambda}_{hol}$ are the electric
and magnetic QCD scales respectively. Their weights will be referred to collectively as $n_\Lambda$. The weight of the Dirac mass
is then constrained to be 
\begin{equation}
n_{m_{D}}=-n_{{\Lambda}}\frac{3N-F}{F}-\frac{N}{F}\,.
\end{equation}
As the three scales have the same modular weights, there can be no relative factors of $T_2$ or $U_2$ between them, and it is natural to assume $\hat{\Lambda}\sim \tilde{\Lambda}_{hol}\sim\Lambda_{hol}$.
For example, if the fields $Q$ and $\tilde{Q}$
are incorporated into the ``no-scale'' structure such that they have weight
$n_{Q}=-1$, then the corresponding modular weights of $\Lambda_{hol}$
and $m_{D}$ are $n_{\Lambda}=-(N+F)/(3N-F)$ and $n_{m_{D}}=1$ respectively. 

Finally the holomorphic magnetic meson is defined as 
\begin{equation}
\Phi=\frac{[Q\tilde{Q}]}{{\Lambda}_{hol}}\,.
\end{equation}
It has weight 
\begin{equation}
n_{\Phi}=n_{\Lambda}\left(\frac{3N-2F}{F}\right)-\frac{F-N}{F}\,.
\end{equation}
Note that the dependence on $n_{\Lambda}$ in eq.(\ref{eq:nqs}) is
proportional to the beta function in the respective theory, and at
fixed points the modular weights of fields are proportional to their
anomaly-free $R$-charges in the global theory. Thus when $F\approx3N/2$ and the magnetic
theory is weakly coupled, $n_{q}\approx n_{\Phi}\approx-1/3$, which can be interpreted as
the appropriate modular weight for them to become free fields at a Gaussian fixed point. 
Likewise the weakly coupled electric theory, when $F\approx3N$, has $n_{Q}\approx-1/3$.
In addition note that a non-zero value for $m_{D}$ breaks both the
anomaly-free $R$-symmetry of the global theory, and the modular symmetry. 

How are these objects related to their physical counterparts? The
physical mass of the quarks is determined by the K\"ahler piece, $K\supset\left(|Q|^{2}+|\tilde{Q}|^{2}\right)(4T_{2}U_{2})^{n_{Q}}$,
so the canonically normalized quark is $\hat{Q}=Q(4T_{2}U_{2})^{n_{Q}/2}$,
while the physical mass is $\hat{m}_{D}=e^{K/2}W_{Q\tilde{Q}}(4T_{2}U_{2})^{-n_{Q}}=m_{D}(4T_{2}U_{2})^{-(n_{Q}+1/2)}$.
Both are modular invariant as they should be.
We must also be careful to distinguish the holomophic scale $\Lambda_{hol}$
from the physical dynamical scale of the theory $\Lambda_{e}$. The
two are related through the gauge thresholds according to eq.(\ref{eq:foo1-1}), which yields
\begin{equation}
\Lambda_{e}=|\Lambda_{hol}|\left(4T_{2}U_{2}\right)^{\frac{n_{\Lambda}}{2}}\, .
\end{equation}
Thus the physical scale $\Lambda_{e}$ can be different
from the holomorphic one, but note that in principle they can be similar
in size, even at large volume: restoring the explicit radii and tilt
dependence, $U_{1}\approx1\implies R_{1}=R_{2}\cos\theta$, and hence
$T_{2}U_{2}\approx R_{2}^{2}-R_{1}^{2}$. One may always choose $R_{2}^{2}\approx R_{1}^{2}+c^{2}$
where $c$ is an $\mathcal{O}(1)$ constant, so that $T_{2}U_{2}\approx c^{2}$.
In this limit the tilt angle is very small, $\sin\theta\approx c/R_{2}$.
This will turn out to be the dynamically relevant limit for the minimisation.

In the large $T_{2}$ and $\tilde{U}_{2}$ limit, eq.(\ref{eq:largevolD})
gives, 
\begin{equation}
\frac{16\pi^{2}}{g_{}^{2}(\mu)}=\frac{16\pi^{2}}{g_{s}^{2}}+\frac{\pi}{3}C_{}\left(T_{2}+\tilde{U}_{2}\right)-b_{}\ln\left(\mu^{2}4T_{2}\tilde{U}_{2}\right),\label{eq:flow-1}
\end{equation}
and hence an approximation for $\Lambda_{e}$,
\begin{equation}
4T_{2}\tilde{U}_{2}\Lambda_{e}^{2}=e^{-\frac{16\pi^{2}}{g_{s}^{2}|b_{}|}+\frac{C_{}}{b_{}}\frac{\pi}{3}\left(T_{2}+\tilde{U}_{2}\right)}\,.\label{eq:rlam-1}
\end{equation}
Since this approximation is valid only in the specific $iU\rightarrow1$ limit,
it is unsurprisingly not modular invariant. Indeed the physical
KK scale is a non-modular invariant quantity, and is given by
the splitting in the spectrum, $M_{KK}=1/\sqrt{4T_{2}\tilde{U}_{2}}$.
As $T_{2}\tilde{U_{2}}\equiv R_{1}^{2}$ it is, unlike $T_{2}U_{2}$,
inevitably large.

The story for the physical magnetic meson is less clear-cut because
it is not possible to determine the normalization precisely. However,
given the modular weight of $\Phi$, it is reasonable to adopt an invariant
K\"ahler potential of (up to irrelevant factors)
\begin{equation}
K\supset|\Phi|^{2}(4T_{2}U_{2})^{n_{\Phi}}+\ldots\label{eq:mesonkahler}
\end{equation}
Thus we work with a normalized field $\hat{\Phi}=\Phi/\gamma$, where
$\gamma\equiv(4T_{2}U_{2})^{-n_{\Phi}/2}$. The canonically normalized field is the modular invariant combination, $\hat{\Phi}=\hat{Q}\hat{\tilde{Q}}/\Lambda_{e}$.
In the free-magnetic window where the ISS mechanism operates,
\begin{equation}
-\frac{1}{3}\leq n_{\Phi}\lesssim1\,,
\end{equation}
with the lower limit corresponding to $2F=3N$.

The aspect of SQCD that we wish to address with this technology is the behaviour
of the soft supersymmetry breaking terms that are induced in the original theory by the SS mechanism. In global theories such
terms can be followed, even through regions of strong coupling, using various
tools, most notably the $R$-current superfield, as described in refs.~\cite{Luty:1999qc,Abel:2011wv}.
For example, properly normalized gaugino masses in the original
SQCD electric theory are mapped to the magnetic dual as 
\begin{equation}
m_{g}^{(mag)} =\frac{2F-3N}{3N-F}m_{g}^{(el)}\,.\label{eq:gaugmap}
\end{equation}
There is a similar (and related) mapping of mass-squared operators
for the squarks and smesons, which in the global theory looks like
\begin{equation}
|\hat{Q}|^{2}+|\hat{\tilde{Q}}|^{2}\rightarrow\left(\frac{2F-3N}{3N-F}\right)\left[|\hat{q}|^{2}+|\hat{\tilde{q}}|^{2}-|\hat{\Phi}|^{2}\right]\,.\label{eq:sq-map}
\end{equation}
These mappings in softly broken {\it global} SQCD theories parametrically
suppress the supersymmetry breaking when the theory is just inside
the free magnetic window $2F\lesssim3N$. 

In a similar fashion, modular symmetry can
track the soft-terms in the effective supergravity theory. Due to its holomorphic nature the gaugino
mass mapping is unchanged. 
 But the mapping for the mass-squareds is {\it different}. Indeed a little work shows
that a generic canonically normalized matter field $\hat{\varphi}$
has soft mass-squared terms
\begin{equation}
m_{\hat{\varphi}}^{2}=m_{3/2}^{2}(1+2n_{\varphi})+\ldots\label{eq:1+2n}
\end{equation}
where the dots indicate loop corrections. Numerical factors in the
normalisation obviously cancel out in the physical mass-squared which depends only on
the modular weights (which is why it was safe to ignore them). {\it In the SQCD supergravity theories, this gives
the following mapping of soft-terms:} 
\begin{align}
m_{\hat{Q}}^{2} & =m_{3/2}^{2}\left[\frac{(3N-F)}{F}n_{\Lambda}+\frac{N}{F}\right]\nonumber \\
m_{\hat{q}}^{2} & =m_{3/2}^{2}\left[\frac{(2F-3N)}{F}n_{\Lambda}+\frac{F-N}{F}\right]\nonumber \\
m_{\hat{\Phi}}^{2} & =m_{3/2}^{2}\left[-\frac{2(2F-3N)}{F}n_{\Lambda}+\frac{2N-F}{F}\right]\,.
\end{align}
One concludes that the relation in eq.(\ref{eq:sq-map}) is not valid in the local theory, but that it {\it would} hold if one were to add a universal $-\frac{1}{3}m_{3/2}^{2}$
constant to all the soft-terms. Combined with the ``1'' in
eq.(\ref{eq:1+2n}), this extra $\frac{2}{3}m_{3/2}^{2}$ contribution
is precisely the gravity mediated piece that is removed by the conformal
compensator technique of ref.~\cite{Luty:1999qc}. Here it is
 a real physical effect, and leads to an interesting
sum-rule,
\begin{equation}
\label{sum-rule}
2m_{\hat{q}}^{2}+m_{\hat{\Phi}}^{2}=m_{3/2}^{2}\,.
\end{equation}
The right-hand side of this equation \textendash{} which would be zero in a global
theory \textendash{} arises entirely from gravity mediation. This sum-rule implies that, in contrast to the global theory,
there is now no choice of parameters  that restores supersymmetry in the magnetic theory.

Eq.(\ref{eq:1+2n}) cannot be the whole story for the scalar
masses: for example no-scale models have massless scalars that have $n_{Q}=-1$.
The additional contribution is of course from the cross-term in $K\supset|Q+\tilde{Q}^{\dagger}|^{2}(4T_{2}U_{2})^{n_{Q}}$.
For models of this form one finds a dimensionful mass-squared operator
in the potential for the canonically normalized fields of the form
\begin{equation}
V_{el}\supset m_{3/2}^{2}(1+n_{Q})|\hat{Q}+\hat{\tilde{Q}}^{\dagger}|^{2}+\ldots\label{eq:1+2n-1}
\end{equation}
The global flavour symmetry is explictly broken as 
\begin{equation}
SU(F)_{L}\times SU(F)_{R}\times U(1)_{B}\times U(1)_{R}\rightarrow SU(F)_{V}\times U(1)_{B}\,,
\end{equation}
by the cross term. All $D$-flat scalar degrees of freedom remain
massless when supersymmetry is spontaneously broken, and imposing
these constraints on the magnetic description (as well as the flavour
symmetry breaking pattern), fixes the magnetic K\"ahler potential to be 
\begin{equation}
K_{mag}\supset|q^{\dagger}+\tilde{q}|^{2}(4T_{2}U_{2})^{n_{q}}+|\Phi^{\dagger}+\Phi|^{2}(4T_{2}U_{2})^{n_{\Phi}}\,,
\end{equation}
with the anti-hermitian part of $\Phi$ remaining massless, but the hermitian and trace parts picking up a mass of order $m_{3/2}^2$. This gives
soft-terms of the form 
\begin{equation}
V_{mag}\supset m_{3/2}^{2}(1+n_{q})|\hat{q}^{\dagger}+\hat{\tilde{q}}|^{2}+m_{3/2}^{2}(1+n_{\Phi})|\hat{\Phi}^{\dagger}+\hat{\Phi}|^{2}\,,
\end{equation}
up to normalisation factors that are irrelevant to the physical masses.   

Finally {with the above information to hand it is possible to check that the relevant physical
processes respect the modular symmetry.} For example a superpotential
can be written for the canonically normalized fields of the effective global theory: 
\begin{equation}
\hat{W}(\hat{\Phi},\hat{q},\hat{\tilde{q}})=We^{-\langle K\rangle/2}=h \hat{\Phi}\hat{q}\hat{\tilde{q}}-\hat{m}_{D}\Lambda_{e}\hat{\Phi}\,,
\end{equation}
where $h=\Lambda_{hol}/\hat{\Lambda}$ is a modular invariant coupling. The conclusion is that the typical induced physical mass scale in
the ISS minimum is $\hat{\mu}=\sqrt{\hat{m}_{D}\Lambda_{e}/h}$.  

Likewise consider the tunneling action in the ISS sector (ignoring the additional soft-terms for $\Phi$ when $n_\Phi\neq -1$). Defining $\varepsilon_{hol}=\sqrt{m_{D}/\Lambda_{hol}}$ and setting $h=1$,
the VEV of the true supersymmetric minima in ISS is determined exactly:
\begin{equation}
\Phi_{0}=\mu_{hol}\varepsilon^{(2F-3N)/2N},
\end{equation}
where $\mu_{hol}^{2}=m_{D}\Lambda_{hol}$. An estimate for the tunnelling
action that takes into account both the factor $e^{K}$ and the normalization
of $\hat{\Phi}$ is then \cite{Intriligator:2006dd}
\begin{equation}
S_{E}\sim2\pi^{2}N\varepsilon_{hol}^{4(2F-3N)/N}(T_{2}U_{2})^{1+3n_{\Phi}}\,.
\end{equation}
Upon inspection, this expression is
the only possible modular invariant combination with the correct functional
dependence on $\varepsilon$ (and this could have been used as a short-cut to derive it). Indeed expressing holomorphic parameters
in terms of physical ones, gives simply 
\begin{equation}
S_{E}\sim2\pi^{2}N\varepsilon^{4(2F-3N)/N}\,,
\end{equation}
where $\varepsilon=\sqrt{\hat{m}_{D}/\Lambda_{e}}$.

\subsection{On $\Lambda_e > M_{KK}$}

An important point for the minimisation is that thanks to the remaining congruence subgroup symmetry there is no longer any reason to prevent $\Lambda_e > M_{KK}$. In particular the matching governed by eqs.(\ref{mag0})-(\ref{mat2}) is 
still valid in these regions of parameter space as long as one bears in mind that the matching is between the effective 4D theories with KK modes integrated out. It is effectively being done at the scale $M_{KK}$. This fact will allow us to avoid the upper constraint in eq.(\ref{eq:hierarchy}).

Let us comment on this more explicitly. The picture of interest is where the original SQCD becomes strongly coupled at 
an energy scale $\Lambda_e > M_{KK}$, when it still contains many light KK 
modes. The effective 4D field theory description at this scale would resemble a truncated 6D theory, while the magnetic theory will be some unknown dual description. The physics of this full theory will be quite messy, so let us see what happens in a toy-model: motivated by the fact that the extra KK states in the spectrum of the electric theory include additional massive KK quarks with Dirac mass terms similar to those in eq.(\ref{mag0}), as per Section~\ref{section2}, consider including just these extra states as a set of $\Delta F$ flavours
with mass $m_{\Delta F}$. One can ``integrate in'' these quarks to find a theory with dynamical scale $\Lambda'_{hol}$ and beta function coefficient $b'=b+\Delta F$. The scale $\Lambda'_{hol}$ would then be regarded as the scale for the truncated 6D theory with its additional $\Delta F$ quarks, and its relation to $\Lambda_{hol}$ can be found by holomorphic matching at the scale $m_{\Delta F}$:
\begin{equation} 
\label{po1}
\left( \frac{\Lambda_{hol}}{m_{\Delta F}}\right)^b=
\left( \frac{\Lambda'_{hol}}{m_{\Delta F}}\right)^{b'}\,.
\end{equation}
The magnetic equivalent of this situation is very well known:  the $m_{\Delta F}$ operator gives rise to a linear meson term that via eq.(\ref{mag1}) induces a Higgsing for the magnetic theory of $\langle q\cdot \tilde{q}\rangle = \hat{\Lambda} m_{\Delta F}$. Hence the ``integrating in'' of the electric theory, corresponds in the magnetic theory to an ``unHiggsing'' from $SU(N)$ to $SU(N+\Delta F)$, which gives a new beta function coefficient $\tilde{b}'=\tilde{b} -2\Delta F$, and an accompanying matching equation
\begin{equation} 
\label{po2}
\left( \frac{\tilde{\Lambda}_{hol}}{\sqrt{\hat{\Lambda}m_{\Delta F}}}\right)^{\tilde b}=
\left( \frac{\tilde{\Lambda}'_{hol}}{\sqrt{\hat{\Lambda}m_{\Delta F}}}\right)^{{\tilde b}'}\, .
\end{equation}
Now, upon inserting eqs.(\ref{po1}),(\ref{po2}), one finds that the 4D matching in eq.(\ref{mat1}) derives from the matching equation of the full theory, namely
\begin{equation}
\label{mat1primed}
\Lambda_{hol}^{\prime -b'_{}}\tilde{\Lambda}_{hol}^{\prime-\tilde{b}'_{}}\sim\hat{\Lambda}^{-(F+\Delta F)}\, .
\end{equation}
The point of this simple exercise is to demonstrate that no explicit powers of $T_2U_2$ can enter when one integrates out modes between $M_{KK}$ and $\Lambda_e$, because that would be in   
violation of the modular symmetry. In principle volume factors could have entered in a modular invariant way via the dependence on 
$\Delta F\sim (\Lambda_e/M_{KK})^d$, but this would have introduced extra powers of $\Lambda$, and  it would also have made the relation between the 6D and 4D dynamical scales singular in the decompactification limit.  
We conclude that the effective 4D relation in eq.(\ref{mat1}) derives from 
the matching relation in the toy-model with all KK modes present in eq.(\ref{mat1primed}), and neither version of the matching can contain factors of $T_2$ or $U_2$.

One does not expect that this conclusion would change if one were to start with the 
full 6D theory truncated at $\Lambda_e$, and its much more complicated magnetic dual (whatever form that may take). Thanks to the modular invariance, the ``integrated in'' 6D relation and the ``integrated out'' 4D relation are  equally valid, although the 4D one is obviously the convenient choice. While it would be interesting to investigate how the 4D duality is embedded in the truncated 6D theory, knowledge of this is not required for the mechanism at hand.  
In particular, $\Lambda_{hol}$ is indeed just
a parameter that specifies the dynamical scale of the effective 4D field theory when one integrates out all the KK physics, and    
$\tilde{\Lambda}_{hol}$ is the relevant dynamical scale for the 4D theory that emerges below $M_{KK}$, regardless of the relative size of $\Lambda_{hol}$ and $M_{KK}$. Note that, if the couplings (i.e. $h=\Lambda_{hol}/\hat{\Lambda}$ and friends) are of order unity, the dynamical scales of the truncated 6D theory are inevitably similar to those of the 4D theory regardless of the volume.

\section{Stabilisation in the string-embedded theories}

\label{seciv}
\subsection{Generic Casimir energy}

With all the necessary ingredients to hand, the minimisation can now be revisited. The generic case is treated in this subsection. The following subsection considers the UV-Casimir case.

To start with, one can deduce from eq.(\ref{sum-rule}) that there are always {\it some} mass-squareds of order $m_{3/2}^{2}$ in the infra-red of the ISS theory. 
It is convenient for the discussion in this and the following subsection to specialise to the weakly coupled case, and take $F\approx3N/2$ giving $n_{q},n_{\tilde{q}},n_{\Phi}\approx-1/3$. This yields positive (physical) mass-squared operators of $\frac{2}{3}m_{3/2}^{2}$, regardless of $n_Q$ and $n_\Lambda$. Generalisation would be straightforward. 

Therefore for the ISS mechanism to work
as before (in particular for the rank-condition to be unchanged) one requires only that $\hat{\mu}^{2}=\hat{m}_{D}\Lambda_{e}/h\gtrsim\frac{2}{3}m_{3/2}^{2}$. There is now the additional attractive feature that gravity mediated contributions act to stabilise the smeson fields around the origin, quenching tunnelling completely. This means one is 
able to relax the conditions in eq.(\ref{eq:hierarchy}): one may work with $\Lambda_e > m_{3/2}\sim M_{KK}$ which then guarantees that 
$\hat{m}_D < \Lambda_e$ ensuring that the physical states all still remain in the ISS theory. It should be stressed that this does {{\it not}} cause a problem for the proper functioning of the ISS mechanism. As discussed in the previous section, the matching of the zero-mode SQCD theories can be done at the scale $M_{KK}$ and goes through as before regardless of the presence of heavier bulk modes. The magnetic ISS phase and the soft-terms all emerge below $M_{KK}$ with $\Lambda_e$ being the appropriate 4D SQCD scale, regardless of the relative size of $\Lambda_e$ and $M_{KK}$, and regardless of what additional states or physics might appear above the KK scale. 

A possible generic difficulty with $\Lambda_e > m_{3/2}$  is rather that the ISS physics could change the original Casimir energy. One can see this sensitivity explicitly, by for example just removing the KK modes below the scale $\Lambda_e$ with an IR cut-off $\tau_2 < 1/\Lambda_e ^2$ on the Schwinger integral: this adds a term that dominates the contribution from the SQCD sector when $\Lambda_e >  M_{KK}$. One can then see the advantage of the UV-Casimir theories whose cosmological constant is unchanged by such a cut-off: they automatically have a Casimir energy that is completely shielded from all IR physics. We take advantage of this feature in the next subsection. By contrast, for the generic case one must assume that the contribution to the cosmological constant from the ISS sector is swamped by the contribution from all the other massless degrees of freedom in the theory, that is $N_{f_{}}^{(0)}-N_{b_{}}^{(0)}\gg N_{f_{ISS}}^{(0)}-N_{b_{ISS}}^{(0)}$. Given the large number of states, this assumption is reasonable.  

To perform the minimisation, let us consider the case $n_Q=-1$, which recall gives $n_{m_D}=+1$. (It is simple but not particularly instructive to generalise.) 
The physical Dirac mass then has the form $\hat{m}_{D}={\alpha_{D}}{\sqrt{4T_{2}U_{2}}}$
where $\alpha_{D}$ is a continuous parameter that must have weight $+1$. 
(Therefore $\alpha_D$ represents an explicit breaking of the modular symmetry much like the 
Dirac mass in the original ISS scheme is an explicit breaking of the anomaly-free $R$-symmetry.) Note that $\alpha_D$ has mass dimension 2: {\it henceforth all dimensionful quantities are in units of $M_s$}. It will become clear that the above choice is 
consistent with the Dirac mass-term being a free parameter in the superymmetric theory. 

Then
using eq.(\ref{eq:e3approx}) we have 
\begin{align}
V & =V_{C}+V_{ISS}\nonumber \\
 & =\frac{\pi^{3}}{120}(N_{f}^{0}-N_{b}^{0})\frac{U_{2}^{3}}{T_{2}^{2}}+N\hat{m}_{D}^{2}\Lambda_{e}^{2}\,,\nonumber \\
 & =\frac{\pi^{3}}{120}\frac{(N_{f}^{0}-N_{b}^{0})}{T_{2}^{2}\tilde{U}_{2}^{3}}+\frac{T_{2}}{\tilde{U}_{2}}\,4N\alpha_{D}^{2}e^{-\frac{16\pi^{2}}{g_{s}^{2}|b_{}|}+\frac{C_{}}{b_{}}\frac{\pi}{3}\left(T_{2}+\tilde{U}_{2}\right)}\,.
\end{align}
Note that  strictly speaking one should add the {\it superpotential} terms corresponding to the two sources of spontaneous supersymmetry breaking 
rather than the vacuum energies, and evaluate the resulting cosmological constant in the full supergravity theory. However the terms in the superpotential comprise a $U$ dependent part from the Casimir energy, and a $\Phi$ dependent part from the ISS 
contribution. The terms that are being neglected by not performing a full treatment can only arise from additional $U-\Phi$ mixing terms in the K\"ahler metric (since $F_U$ and $F_\Phi$ are the only non-zero $F$-terms); by flavour symmetry these have extra factors of $\langle{\Phi}\rangle$ which are zero at leading order. 

The minimisation conditions give 
\begin{equation}
\tilde{U}_{2}=\frac{3}{2}T_{2}+\frac{15b_{}}{2\pi C_{}}\,.
\end{equation}
Assuming that the volume ends up at $T_{2}\gg1$ (as will be verified
in a moment), one may neglect the second term and use \begin{equation}T_{2}U_{2}=\frac{2}{3}+\mathcal{O}(1/T_{2})\, , \end{equation} and hence $M_{KK}\approx\sqrt{2/3}\,\,T^{-1}_2$.
Note that $\hat{m}_D\approx \sqrt{8/3} \,\alpha_D$ regardless of the eventual scale of supersymmetry breaking. Therefore $\alpha_D$ can indeed be considered to be 
a parameter of the supersymmetric theory.

\begin{figure}
\noindent \begin{centering}
~\\
\qquad$\log_{10}V(T_{2},\tilde{U}_{2})\,\,\,\,\,\,\,\,\,\,\,\,\,\,\,\,\,\,\,\,\,\,\,\,\,\,\,\,\,\,\,\,\,\,\,\,\,\,\,\,\,\,\,\,\,\,\,\,\,\,\,\,\,\,\,\,\,\,\,\,\,\,\,\,\,\,\,\,\,\,\,\,\,\,\,\qquad\qquad \log_{10}V(T_{2},\tilde{U}_{2})\,\,\,\,\,\,\,\,\,\,\,\,\,\,\,\,$
\par\end{centering}
\noindent \centering{}
\includegraphics[bb=0bp 0bp 400bp 380bp,clip,scale=0.55]{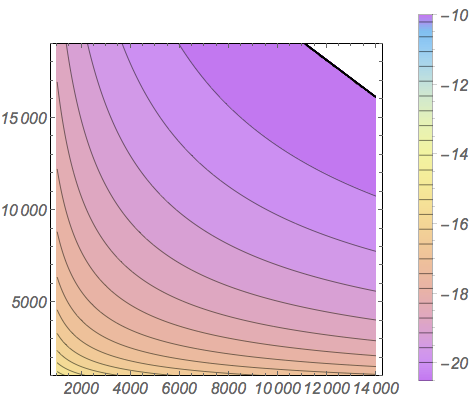}
\includegraphics[bb=0bp 0bp 470bp 380bp,clip,scale=0.55]{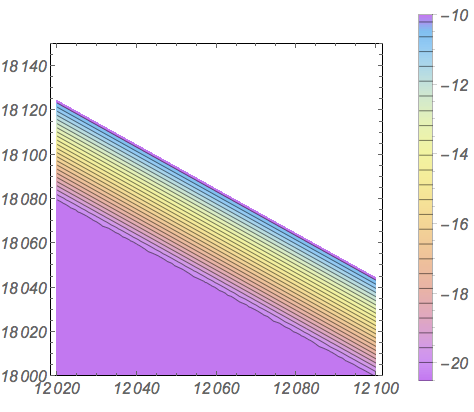}\\
\caption{\label{fig:E3-2}The potential for $b_{}=-100$, $C_{}=-50$,
$(N_{f}^{0}-N_{b}^{0})=N=10$, $\alpha_{D}=10^{-4}$ and $g_{s}=10^{-2}$.
The approximation in eq.(\ref{eq:minimar}) gives a minimum at $T_{2},\tilde{U}_{2}=12064,\,18096$ (in string units) 
respectively. As in the 5D case, the minimum lies close to a boundary
in moduli space beyond which the effective field theory theory description breaks down as the dynamical QCD scale exceeds the string scale. The dynamical scale is $\Lambda_{e}\approx 35 M_{KK}$ but it can be easily brought closer to $M_{KK}$ with different choices of parameters, while $\Lambda_{e}\gg \hat{m}_{D}$ over the whole parameter space.}
\end{figure}

The potential becomes 
\begin{equation}
V(T_{2})=\frac{\pi^{3}}{405}\frac{(N_{f}^{0}-N_{b}^{0})}{T_{2}^{5}}+\frac{8N\alpha_{D}^{2}}{3}e^{-\frac{16\pi^{2}}{g_{s}^{2}|b_{}|}+\frac{C_{}}{b_{}}\frac{5\pi}{6}T_{2}} \,.
\end{equation}
The remaining one-dimensional minimisation can be done analogously
to that in the 5D model of Section 2. Using that notation,
the fiducial scale of eq.(\ref{eq:g2}) and the $T_{2}$ VEV are given
by 
\begin{align}
{\mu_{0}^{2}} & \approx\left(\frac{5\pi C_{}}{6b_{}}\right)^{5}\frac{\pi^{3}}{6^9}\frac{(N_{f}^{0}-N_{b}^{0})}{N\alpha_{D}^{2}}\,\nonumber \\
& \approx \frac{4\times 10^{-4}}{\alpha_{D}^{2}} \left(\frac{C_{}}{b_{}}\right)^{5}\frac{(N_{f}^{0}-N_{b}^{0})}{N}\,\, ,\nonumber \\
\frac{5\pi C_{}}{6b_{}}T_{2} & \approx\frac{16\pi^{2}}{|b_{}|g_{s}^{2}}+\ln {\mu_{0}^{2}}\,\,,\label{eq:minimar}
\end{align}
where, recall, the dynamical scale is then given by $\frac{\Lambda_{e}}{M_{KK}}={\mu_{0}}$ (in string units).
As mentioned above, with maximal SS phases, in order to avoid the SS soft-terms interfering with the ISS mechanism we choose $\Lambda_{e}\gtrsim M_{KK}$. From the above, assuming $(N_{f}^{0}-N_{b}^{0})\sim N$
and $C_{}\sim b_{}$ requires $\alpha_{D}^{2}\ll 1$, which is consistent with $\hat{m}_D\ll M_s$. Indeed restoring the string scale we have 
\begin{equation}
\frac{\Lambda_e}{M_{KK}}\approx  \sqrt{
{ 10^{-3}} \left(\frac{C_{}}{b_{}}\right)^{5}
\frac{(N_{f}^{0}-N_{b}^{0})}{N}
}\times \frac{M_s}{\hat{m}_D}\, .
\end{equation}
Summarising the 6D case then, when $g_{s}\ll1$,
the minimum is at 
\begin{equation}
T_{2}\approx\frac{2}{3U_{2}}\approx\sqrt{\frac{2}{3} }M^{-1}_{KK}\approx\frac{96\pi}{5|C_{}|g_{s}^{2}}\,,
\end{equation}
with $\Lambda_{e}\gtrsim M_{KK}.$ As in the 5D case the potential rises exponentially fast beyond the minimum until 
$\Lambda_e$ surpasses $M_s$. A numerical example is shown in fig.~(\ref{fig:E3-2}).

\subsection{UV-Casimir energy balanced against a gaugino condensate}

Next we consider the $N_f^0=N_b^0$ theories. As discussed earlier the Casimir energy in these cases is generated entirely by UV modes, 
so it is completely insensitive to the low energy physics. This separation 
is very interesting in the current context of balancing competing Scherk-Schwarz induced terms against non-perturbative IR physics, because it 
suggests that whatever mechanism is devised will be very robust. Moreover the two contributions to the cosmological constant may be consistently determined 
independently even though they necessarily involve the same moduli. In terms of the Schwinger integral, one can envisage the integrand 
as having two separate peaks, one at the stringy UV end and the other at the non-perturbative IR end. Therefore, one may simply add the two terms, which will be referred to as $V_{UV}$ and $V_{IR}$, in the cosmological constant. Indeed $V_{UV}$ is computed in the string theory, while $V_{IR}$ can be computed independently in the low energy effective field theory. 

This opens up possibilities for stabilisation with non-perturbative physics that would otherwise be rather difficult to treat. For example 
gaugino condensation is now an attractive option for our IR physics rather than the ISS mechanism. Note that by contrast a {\it standard}\, SS Casimir energy (as considered in the previous section) balancing against a gaugino condensate would require a treatment of both terms simultaneously because they are functions only of $S,T,U$ and are not independent; essentially everything in that case would be happening in the IR, so it would be necessary to determine the full one-loop effective supergravity theory in order to compute the cosmological constant.

To see this in practice, consider a single gaugino condensation contribution to $V_{IR}$. The minimisation will now be done with all three fields, properly including the dynamics of the dilaton $S$ itself. However the philosophy is the same, namely we expect to end up in a stable or metastable minimum that has relatively large $S$ compensated by relatively large $T$ and $\tilde{U}$. 

The IR contribution to the potential is calculated in supergravity, incorporating the superpotential $W_{SS}$ in eq.(\ref{eq:shtom}) for the Scherk-Schwarz background, in addition 
to the gaugino condensate, that is $W_{IR}=W_{SS}+W_{gc}$.  The latter is described by the well-known superpotential  
\begin{eqnarray}
W_{gc}=d \Lambda_{hol}^3\, ,
\end{eqnarray}
where $d$ is a constant, and now 
\begin{equation}\Lambda_{hol} \approx e^{-\frac{8\pi^2}{|b|}S + \frac{C}{b}\frac{\pi}{6}(T+\tilde{U} )}\end{equation} 
is the holomorphic scale for the  
pure Yang-Mills theory.  Eq.(\ref{eq:nlrel}) with $F=0$ and $b_0=-3N$ correctly gives $n_{W_{gc}}=-1$.
 The approximation refers to $T_2,\, \tilde{U}_2\gg 1$ near $iU=1$ and as discussed earlier it breaks the 
modular symmetries. In the Scherk-Schwarz background, adding $W_{SS}$  then incorporates the effect of the shifted mass 
spectrum. We know that the potential without $W_{gc}$ is entirely flat so one can anticipate that the resulting contribution 
involves powers of $W_{gc}$. 

Some care is required regarding phases: bearing in mind the cosmological constant discussion in Section \ref{sec:casi}, one can anticipate that 
$U_1$ and $T_1$ will ultimately be fixed to zero by $V_{UV}$, and therefore one does not need to consider them further.  However the phase of the dilaton $S_1$ remains as a free field that is fixed by the gaugino condensate. 
 
Using eq.(\ref{eq:master-exp}), the potential is conveniently arranged (at $U_1=T_1=0$) as   
\begin{eqnarray}
V ~=~ V_{UV}+V_{IR} &=&  2{(N_{f}^{1}-N_{b}^{1})}\,T^{-3/4}_{2}\tilde{U}^{-7/4}_{2}   e^{-2\pi \sqrt{T_2 \tilde{U}_2}  }
+ B \left(| \Lambda^3_{hol}| - \frac{A}{B} \right)^2 - \frac{|A|^2}{B} \nonumber \\
A\,\frac{S_2T_2}{d} & = & \frac{1}{2\sqrt{2}}  \left( 1 +\log|\Lambda^3_{hol}| \right)  \\
B\,\frac{S_2T_2}{d^2\tilde{U}_2} & = & \frac{1}{2} \log|\Lambda^3_{hol}|(\log|\Lambda^3_{hol}|-1) -\frac{\pi}{2} \frac{C}{b}\left((T_2+\tilde{U}_2) \log|\Lambda^3_{hol}|-\tilde{U}_2\right) 
+\left(\frac{\pi}{2}\frac{C}{b}\right)^2 (T_2^2+T_2\tilde{U}_2+\tilde{U}_2^2 )\, . \nonumber
\end{eqnarray}
The entire $S_2$ dependence is contained within the $e^K$ prefactors and the $|\Lambda_{hol}|$ dependence, while $S_1$ simply adjusts the phase of $\Lambda_{hol}$ so that it comes to rest where it minimises the square with a relative minus sign as shown. The minimisation with respect to the dilaton is then dominated by the complete square term, which gives the approximation 
\begin{eqnarray} |\Lambda_{hol}| &\approx &
 A/B\left(1 + {\cal{O}}(24\pi^2 S_2/|b|)  \right)\nonumber \\
 &\approx & \frac{1}{2\sqrt{2}d} \left(\frac{\pi}{2} \frac{C }{b}\right)^{-2} \frac{1}{\tilde{U}_2(T_2^2+T_2\tilde{U}_2+\tilde{U}_2^2 )}  
  \, . \end{eqnarray} 
The error on the right hand side of this equation is due to the $e^K$ pre-factor and is negligible when the gauge coupling at the string scale ends up being weak (as is the case of interest).  The $A/B$ term on the right hand side depends only logarithmically on $\Lambda_{hol}$; the approximation can be improved by iteration if required but as long as the volume $T_2$ is large, the zeroth order expression shown on the second line is sufficiently accurate. 

The potential is qualitatively different from that in the ISS case because the single gaugino condensate does not by itself give a minimum in $T_2$ or $U_2$. In fact 
without the $V_{UV}$ contribution the potential has a runaway to small moduli (where our approximations break down) or to infinity. With $V_{UV}$ however 
a minimum is found where the two terms $V_{UV}$ and $V_{IR}$ balance, giving rise to the novel phenomenon that the non-perturbative low-energy contribution self-tunes to be of the same order as the exponentially suppressed UV-Casimir energy. A framework in which an exponentially small UV cosmological constant governs and stabilises non-perturbative IR physics without being disrupted itself seems of general interest.

\begin{figure}
\noindent \begin{centering}
~\\
$V(T_{2},\tilde{U}_{2})\qquad\,\,\,\,\,\,\,\,\,\,\,\,\,\,\,\,\,\,$
\vspace{0.2cm}
\par\end{centering}
\noindent \centering{}
\includegraphics[bb=0bp 0bp 520bp 416bp,clip,scale=0.55]{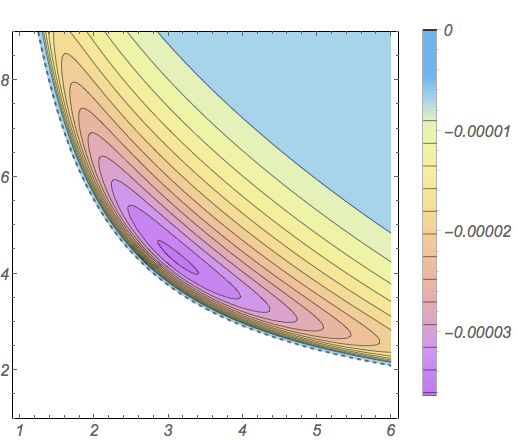}\\
\caption{\label{CasiPot}The potential for a single gaugino condensate in a Scherk-Schwarz background, for $b_{}=-80$, $C=-20$,
$(N_{f}^{1}-N_{b}^{1})=10^6$ with $d=0.1$. The dashed line marks where the IR contribution to the potential is equal to the UV one. 
In practice the pre-factors make very 
little difference to the qualitative form of the potential, but move the minimum along $T_2\approx \frac{2}{3}\tilde{U}_2$. }
\end{figure}

An example potential is shown in fig.~\ref{CasiPot} for a typical set of parameters. In addition the plot shows the line where $V_{UV}=V_{IR}$ close to the actual minimum. The nett result is a minimum in which all the moduli are stabilised and $\Lambda_{hol} \sim  M_s/10$.    
Notice that the approximation $T_2\approx \frac{2}{3} \tilde{U}_2$ at the minimum still holds. This example 
takes $N_{f}^{1}-N_{b}^{1}=10^6$ which may seem large, but one should recall that there are very many excitations at the first string excitation level, and in fact this number is quite typical. Not surprisingly, reducing this number (and increasing $d$) moves the minimum closer to the origin, where neglected contributions to $V_{UV}$ such as those from winding modes will start to become important. Further discussion of the latter along with explicit examples can be found in the recent work of ref.~\cite{Florakis:2016ani}, and it would be interesting to incorporate these additional terms in detail. 

\section{Conclusion}

In summary, it is argued that a general means of addressing the decompactification problem dynamically is to balance 
non-perturbative physics contributions to the vacuum energy against the Casimir energy in Scherk-Schwarzed theories. Due to universality in both the threshold corrections and the gauge couplings, the stable minimum will have consistently large (order one) gauge couplings for any gauge group that shares the same ${\cal N}=2$ 
beta function for bulk modes as the gauge group taking part in the minimisation. By contrast gauge symmetries with the wrong-sign beta function 
will remain as effectively global symmetries. 

Both the ISS mechanism and a single gaugino condensate were considered for the stabilising non-perturbative physics in the case of compactification from 6D to 4D in heterotic strings. In either case, both the Scherk-Schwarz contribution and the non-perturbative contribution to supersymmetry breaking can be written as superpotential terms in $\mathcal{N}=1$ theories, which spontaneously break supersymmetry.

The ISS mechanism is interesting because it gives novel cross-checks based on the residual modular symmetry of the theory, and also allows one to handle the supersymmetry breaking from the ISS mechanism and the Scherk-Schwarz breaking simultaneously. By contrast the gaugino condensate is interesting when the original Scherk-Schwarzed theory retains Bose-Fermi degeneracy and has exponentially suppressed cosmological constant. 
An important aspect of the SS induced cosmological constant in this case is that it is entirely generated by heavy modes and as such is completely immune to any non-perturbative physics that might be added in the IR to provide a balancing contribution. 
It allows very simple treatment of the minimisation which in this case takes place at moderate volume. A full treatment in this generic set-up (that is, including the stabilisation of the compactification moduli as well as the original dilaton) was presented. The energetic separation between competing and balancing UV and IR induced terms in the potential makes stabilisation very robust, and seems to be something that has not been remarked upon before. It would be of interest to apply the mechanism to explicit examples, such as the models discussed recently in ref.~\cite{Florakis:2016ani}, which has some intriguing overlaps with the work described here.\\ \mbox{ }\vspace{1cm} \\

\noindent {\bf Acknowledgements:} I would like to thank Keith Dienes, Costas Kounnas, Alberto Mariotti, Eirini Mavroudi and Carlos Tamarit for discussions. 

\newpage 

\appendix
\section{The $SL(2,\mathbb{Z})_U$ and $SL(2,\mathbb{Z})_U$ modular symmetries}

The heterotic modular symmetries begin life
as subgroups of the exact $O(16+d,d,\mathbb{Z})$ target-space automorphisms
of the Narain lattice \cite{Giveon:1994fu,LopesCardoso:1994is,Antoniadis:1994hg}. The transformations 
under $SL(2,\mathbb{Z})_U$ and $SL(2,\mathbb{Z})_U$ are 
presented here for reference. Under $SL(2,\mathbb{Z})_{T}$, 
the fields transform as 
\begin{eqnarray}
T & \rightarrow & \frac{aT-ib}{icT+d}\, ,\nonumber \\
U & \rightarrow & U-ic\frac{\phi\phi'}{icT+d}\, ,\nonumber \\
S & \rightarrow & S-\delta_{GS}\log(icT+d)\, ,\nonumber \\
\phi,\phi' & \rightarrow & \frac{\phi,\phi'}{icT+d}\, , \label{eq:discrete}
\end{eqnarray}
with $a,b,c,d\in\mathbb{Z}$ and $ad-bc=1$, while the $U$-modular
transformation $SL(2,\mathbb{Z})_{U}$ is 
\begin{eqnarray}
U & \rightarrow & \frac{aU-ib}{icU+d}\, ,\nonumber \\
T & \rightarrow & T-ic\frac{\phi\phi'}{icU+d}\, ,\nonumber \\
S & \rightarrow & S-\delta_{GS}\log(icU+d)\, ,\nonumber \\
\phi,\phi' & \rightarrow & \frac{\phi,\phi'}{icU+d}\,.\label{eq:discrete-1}
\end{eqnarray}
Some useful identities under the $iT\rightarrow-1/iT$ transformation
of the $SL(2,\mathbb{Z})_{T}$ modular group for example, are
\begin{eqnarray}
T+\bar{T} & \rightarrow & \frac{T+\bar{T}}{|icT+d|^{2}}\nonumber \\
\eta(iT)^{2} & \rightarrow & \left(icT+d\right)\eta(iT)^{2}\nonumber \\
|\eta(iT)|^{4}\left(T+\bar{T}\right) & \rightarrow & |\eta(iT)|^{4}\left(T+\bar{T}\right)\,,
\end{eqnarray}
so that the K\"ahler potential $
K = -\log\left(4T_{2}U_{2}-|\phi+\bar{\phi'}|^{2}\right)$
transforms as $K\rightarrow K+\log|icT+d|^{2}$. Thus the superpotential
has to have weight $-1$ under $SL(2,\mathbb{Z})_{T,U}$ .

\end{document}